%% file: LCS.tex
\documentclass[letterpaper,11pt]{article}
\usepackage{ltexpprt}
\usepackage{amsmath,amssymb}
\usepackage[usenames, dvipsnames]{color}
\usepackage[normalem]{ulem} 
\usepackage{fullpage}
\usepackage[numbers, sort]{natbib}
\usepackage{algpseudocode}
\usepackage{tikz, pgfplots}
\usepackage{enumerate}
\usepackage{hyperref}
\usepackage{xspace,color}
\usepackage{graphicx}
\usepackage{caption}
\usepackage{subcaption}
\usepackage{enumitem,linegoal}
\usepackage[linesnumbered,ruled,vlined,algo2e]{algorithm2e}
\usepackage{comment}	
\usepackage{ctable}

\newtheorem{observation}{Observation}[section]
\usepackage{mathtools}
\usepackage[flushleft]{threeparttable}
\definecolor{refkey}{rgb}{0,0,1}
\definecolor{labelkey}{rgb}{1,0,0}
\usepackage{hyperref}
\hypersetup{
	colorlinks   = true,
	citecolor    = gray
}

\usepackage{tocloft}

\renewcommand{\cftsubsecfont}{\normalfont\hypersetup{linkcolor=orange}}
\renewcommand{\cftsubsecafterpnum}{\hypersetup{linkcolor=green}}

\usepackage{bbm}
\usepackage{tcolorbox}

\usepackage{footnote}
\makesavenoteenv{tabular}
\makesavenoteenv{table}

\definecolor{crimsonglory}{rgb}{0,0,0}

\newtheorem{definition}{Definition}
\newtheorem{problem}{Problem}

\newtheorem{thm}{Theorem}

\makeatletter
\def\GrabProofArgument[#1]{ #1: \egroup\ignorespaces}
\def\proof{\noindent\textbf\bgroup Proof%
	\@ifnextchar[{\GrabProofArgument}{. \egroup\ignorespaces}}

\makeatother

\usetikzlibrary{arrows,shapes,snakes,automata,backgrounds,petri,calc,decorations.markings}
\usepackage[latin1]{inputenc}

\newcounter{proccnt}

\newcommand{\konote}[1]{}

\usepackage[normalem]{ulem} 

\begin{document}

\input{macros.tex}

\title{Approximating \lcs\ in Linear Time:\\ Beating the $\sqrt{n}$ Barrier\thanks{A portion of this work was completed while some of the authors were visiting Simons Institute for Theory of Computing.} \thanks{Supported in part by NSF CAREER award CCF-1053605, NSF AF:Medium grant CCF-1161365, NSF BIGDATA grant IIS-1546108, NSF SPX grant CCF-1822738, and two small UMD AI in Business and Society Seed Grant and UMD Year of Data Science Program Grant}}
\author{MohammadTaghi Hajiaghayi\thanks{University of Maryland}\\
\and
Masoud Seddighin\thanks{Sharif University of Technology}\\
\and
Saeed Seddighin\footnotemark[2]\\
\and
Xiaorui Sun\thanks{University of Illinois at Chicago}\\
}
\date{}

\maketitle


\fancyfoot[R]{\scriptsize{Copyright \textcopyright\ 2019 by SIAM\\
Unauthorized reproduction of this article is prohibited}}





\begin{abstract} \small\baselineskip=9pt 
\input{abstract}
\end{abstract}

\input{introduction}
\input{preliminaries}

\input{organization}
\input{our-contribution.tex}

\input{formal}
\input{BoundingSolutionSize}
\input{BoundingFrequencies}

\input{Boundingsumfreq}
\input{acknowledgement}

\bibliographystyle{abbrv}	
\bibliography{lcs}

\end{document}

%% file: macros.tex
\renewcommand{\bar}[1]{\mkern 1mu\overline{\mkern-1mu#1\mkern-1mu}\mkern 1mu}
\newcommand{\slow}{s_{\mathsf{low}}}
\newcommand{\barslow}{\bar{s}_{\mathsf{low}}}

\newcommand{\cp}{\mathsf{cp}}
\newcommand{\mask}[1]{\mathsf{mask}(#1)}
\newcommand{\opt}{\mathsf{opt}}
\newcommand{\LLL}{\mathsf{L}}
\newcommand{\HHH}{\mathsf{H}}

\newcommand{\shigh}{s_{\mathsf{high}}}
\newcommand{\barhigh}{\bar{s}_{\mathsf{high}}}
\newcommand{\fr}{\mathsf{fr}}

\newcommand{\exponent}{0.497956}

\newcommand{\sbar}{\bar{s}}
\newcommand{\dom}{\textsf{dom}}

\newcommand{\tildbar}{\bar{O}}
\newcommand{\tildo}{\bar{o}}
\newcommand{\omegabar}{\bar{\Omega}}
\newcommand{\tildomega}{\bar{\omega}}
\newcommand{\tildorder}{\widetilde O}

\newcommand{\boundingthesize}{\textsf{bounding the solution size}}

\newcommand{\cmax}{\mathsf{c^{max}}}
\newcommand{\first}{\mathsf{first}}
\newcommand{\lcs}{\textsf{LCS}}
\newcommand{\llcs}{\textsf{lcs}}
\newcommand{\clcs}{\widetilde{\textsf{lcs}}}
\newcommand{\flcs}{\llcs(s,\bar{s})}
\newcommand{\solution}{t}
\newcommand{\rarusblocks}{\widetilde{U}_2}
\newcommand{\crebrisblocks}{\widetilde{U}_1}
\newcommand{\block}{b}
\newcommand{\barblock}{\bar{b}}
\newcommand{\Usize}{n^{1/2-26\delta -4\zeta-\eta}}
\newcommand{\RandUsize}{n^{1/6-65/3\delta-10/3\zeta-4/3\eta}}
\newcommand{\brruntime}{n^{1/2 - 1/37 + (221/37)\delta+(10/37)\eta}}
\newcommand{\stringname}{sequence}
\newcommand{\blocktobloc}{\textsf{block to block}}
\newcommand{\cntchar}{\textsf{counting the characters}}
\newcommand{\samarr}{\textsf{sampled array}}
\newcommand{\rash}{\textsf{random shifting}}
\newcommand{\freqbnd}{\textsf{bounding the frequency of each character}}

%% file: abstract.tex
Longest common subsequence (\lcs) is one of the most fundamental problems in combinatorial optimization. Apart from theoretical importance, \lcs\ has enormous applications in bioinformatics, revision control systems, and data comparison programs\footnote{A notable example is the UNIX application \textsf{diff}}. Although a simple dynamic program computes \lcs\ in quadratic time, it has been recently proven that the problem admits a conditional lower bound and may not be solved in truly subquadratic time \cite{abboud2015tight}. In addition to this, \lcs\ is notoriously hard with respect to approximation algorithms. Apart from a trivial sampling technique that obtains a $n^{x}$ approximation solution in time $O(n^{2-2x})$ nothing else is known for \lcs. This is in sharp contrast to its dual problem \textit{edit distance} for which several linear time solutions are obtained in the past two decades~\cite{bar2004approximating,batu2006oblivious,andoni2012approximating,andoni2010polylogarithmic,bestpaper}. 

In this work, we present the first nontrivial algorithm for approximating \lcs\ in linear time. Our main result is a linear time algorithm for the longest common subsequence which has an approximation factor of $O(n^{\exponent})$. This beats the $\sqrt{n}$ barrier for approximating \lcs\ in linear time.

%% file: introduction.tex
\section{Introduction}
Longest common subsequence (\lcs) is a central problem in combinatorial optimization and has been subject to many studies since the 1950s \cite{paterson1994longest, hunt1977fast,bergroth2000survey,ullman1976bounds,hirschberg1975linear,nakatsu1982longest,arslan2005algorithms,hsu1984computing,jiang2000longest,masek1980faster,bille2008fast,grabowski2016new,abboud2015tight,bellman1957dynamic}. In this problem, we are given two strings $s$ and $\bar{s}$ of size $n$ and the goal is to find the longest string $t$ which appears as a (not necessarily consecutive) subsequence in both $s$ and $\bar{s}$. It is known since 1950~\cite{bellman1957dynamic} that \lcs\ can be dynamically solved in time $O(n^2)$ using the following recursive formula:
\begin{equation*}
T[i][j] =
\begin{cases}
1 + T[i-1][j-1], & \text{if }s_i = \bar{s}_j \\
\max \Bigg\{\begin{array}{c}T[i-1][j-1],\\T[i][j-1],\\T[i-1][j]\end{array}\Bigg\}\hspace{0.cm}\color{white}.\color{black} & \text{if }s_i \neq \bar{s}_j.
\end{cases}
\end{equation*}

Although this solution is almost as old as the emergence of dynamic programming as an algorithmic tool, thus far the only improvements to this algorithm were limited to shaving polylogarithmic factors from the running time \cite{masek1980faster,bille2008fast,grabowski2016new}. Such failures have been partially addressed by the recent result of Abboud, Backurs, and Williams \cite{abboud2015tight} wherein the authors show a conditional quadratic lower bound on the computational complexity of any algorithm that computes the exact value of \lcs\ (this is also shown independently by Bringmann and K{\"u}nnemann~\cite{bringmann2015quadratic}).

Indeed, quadratic time is too costly for several applications with large datasets and therefore an interesting question is how best we can approximate \lcs\ in subquadratic and in particular in near linear time.  Unfortunately, with respect to approximation algorithms, also nothing nontrivial is known about \lcs. A trivial observation shows that a sampling technique improves the quadratic time solution of \lcs\ to an $O(n^{2-2x})$ algorithm for \lcs\ with approximation factor $O(n^x)$. In particular, when we restrict the running time to be linear, this gives an algorithm for \lcs\ with approximation factor $O(\sqrt{n})$ which is the only known linear time solution for \lcs. 

In contrast, several breakthroughs have advanced our knowledge of approximation algorithms for edit distance which is seen as the dual of \lcs. Similar to \lcs, edit distance can be solved in quadratic time via a simple dynamic programming technique. Moreover, edit distance admits a conditional lower bound and cannot be solved in truly subquadratic time unless SETH fails~\cite{backurs2015edit}. Perhaps coincidentally, the first linear time algorithm for edit distance also has an approximation factor of $O(\sqrt{n})$~\cite{landau1998incremental}. The seminal work of Bar-Yossef, Jayram, Krauthgamer, and Kumar ~\cite{bar2004approximating} breaks the $\sqrt{n}$ barrier for edit distance by giving a near linear time algorithm with approximation factor $O(n^{0.43})$. Since then, the approximation factor is improved in  series of works   to $O(n^{0.34})$~\cite{batu2006oblivious}, to $O(2^{\tildorder(\sqrt{\log n})})$~\cite{andoni2012approximating} , and to polylogarithmic~\cite{andoni2010polylogarithmic}. A recent work of Boroujeni \textit{et al.}~\cite{boroujeni2018approximating} obtains a  constant approximation \textbf{quantum} algorithm for edit distance that runs in truly subquadratic time. This improvement is achieved by exploiting triangle inequality which holds for edit distance. The quantum element of their algorithm comes from the Grover's search which they use to extract the edges of a sparse graph which represents the areas of the strings with small distances. Later, Chakraborty \textit{et al.}~\cite{bestpaper} turn this into a classic algorithm by replacing the quantum component of the algorithm by an alternative randomized technique to obtain a classic solution. None of these results directly or indirectly imply any algorithm for \lcs.

When the alphabet size is small, a simple counting algorithm obtains a $|\Sigma|$-approximation of \lcs\ in linear time ($|\Sigma|$ is the size of the alphabet). However, efforts to improve upon this simple solution have failed thus far and this led many in the community to believe that an improved algorithm may be impossible. This is also backed by several recent hardness results \cite{abboud2018fast,abboud2017towards}. The small alphabet size setting is particularly interesting since DNAs consist of only four symbols and therefore approximating the \lcs\ of two DNAs falls within this setting. However, even in the case of DNAs, one may be interested in approximating the number of ``blocks of nucleotides" that the two DNAs have in common where each block carries some meaningful information about the genes. In this case, every block can be seen as a symbol of the alphabet and thus the size of the alphabet is large. 

In this work, we beat the $O(\sqrt{n})$ barrier for \lcs\ and present the first nontrivial algorithm for approximating the longest common subsequence problem. Our algorithm runs in linear time and has an approximation factor of $\tilde{O}(n^{\exponent})$. Despite the simplicity of our algorithm, our analysis is based on several nontrivial structural properties of \lcs.

\subsection{Related Work}
For  exact solutions, most of the previous works are focused on the cases that  (i) the number of pairs of positions of the two strings with equal characters is small \cite{hunt1977fast,apostolico1987longest} (ii) the solution size is close to $n$ \cite{nakatsu1982longest,wu1992fast}, or (iii) polylogarithmic improvements in the running time~\cite{masek1980faster,bille2008fast,grabowski2016new}.

On the approximation front, most of the efforts have been focused on the following question: ``can a linear time algorithm approximate \lcs\ within a factor $o(|\Sigma|)$ where $|\Sigma|$ is the number of symbols of the two strings?". Thus far, only negative results for this problem are presented~\cite{abboud2018fast,abboud2017towards}.

In an independent work, Rubinstein \textit{et al.}~\cite{lcdquantum} give an improved approximate solution for \lcs\ that runs in truly subquadratic time.

Very recently, Hajiaghayi \textit{et al.}~\cite{lcsnew} give massively parallel $1+\epsilon$ approximation algorithms for both edit distance and longest common subsequence. Their algorithms run in quadratic time and the round complexity of both algorithms is constant.

%% file: preliminaries.tex
\section{Preliminaries}
Throughout this paper, we study the longest common subsequence problem (\lcs). In this problem, we are given two strings $s$ and $\bar{s}$ of length $n$ and we wish to find/approximate the longest sequence of (non-continuous) matches between the two strings. 

\begin{problem}
Let $s$ and $\bar{s}$ be two strings of length $n$ over an alphabet $\Sigma$. In the $\lcs$ problem, we want to find a string $t$ with the maximum length such that $t$ is a subsequence of both $s$ and $\bar{s}$. In other words, one can obtain $t$ from both $s$ and $\bar{s}$ by removing some of the characters.
\end{problem}

We call each element $c$ of $\Sigma$ a \textit{symbol} and each element of a string a \textit{character}. A symbol may appear several times in a string but a character is associated with a specific position.

Also, we denote the alphabet size by $m = |\Sigma|$ and let $R := |\{(i,j) | s_i = \bar{s}_j\}|$ be the number of pairs of characters of the two strings that have the same symbol. The assumption that both of the strings are of the same size $n$ is w.l.o.g. since one can pad enough dummy characters to the end of either $s$ or $\bar{s}$ to ensure their lengths are equal. For brevity, we use $\lcs$ to refer to the longest common subsequence problem and denote the solution for two strings $s$ and $\bar{s}$ by $\flcs$. For a symbol $c \in \Sigma$, we denote the frequency of $c$ in a string $t$ by $\fr_c(t)$. Furthermore, we denote a substring of a string $t$ by $t[i,j]$ where $i$ and $j$ are the starting and ending points of the substring. Similarly, we denote  the $i$'th character of a string $t$ by $t_i$.

Finally, a string $t$ is an $\alpha$-approximation of $\flcs$, if $t$ is a substring of both $s$ and $\sbar$, and $|t| \geq |\flcs|/\alpha$.

%% file: organization.tex
\section{Organization of the Paper}
Our solution consists of four algorithms (Algorithms \ref{alg:2}, \ref{alg:7}, \ref{alg:3}, and \ref{alg:4}). Algorithms \ref{alg:2} and \ref{alg:7} are simpler both in terms of implementation and analysis. However, the main novelty of our work is the design and analysis of Algorithms \ref{alg:3}, and \ref{alg:4}. 

We first give an overview of our solution in Section \ref{overview} and discuss the ideas of the algorithms. In Section \ref{overview} we simplify the problem by defining a new notation $\bar{O}$ and ignoring the factors that can be hidden inside this notation. Indeed, our analysis in Section \ref{overview} only shows how an approximation factor of $O(n^{1/2-\epsilon})$ is possible for small enough $\epsilon > 0$.

Section \ref{sec:overview:5} gives a formal proof for the approximation factor of our algorithm. This is indeed achieved by rigorous analysis of Algorithms \ref{alg:2}, \ref{alg:7}, \ref{alg:3}, and \ref{alg:4}. The proofs of Section \ref{bbrs} are substantially more complicated that the ones we bring in Section \ref{overview} as in Section \ref{overview} we only resort to obtaining an $O(n^{1/2-\epsilon})$ solution for some $\epsilon > 0$. However, in Section \ref{bbrs} the goal is to minimize the exponent of the approximation factor.

%% file: our-contribution.tex
\section{An Overview of the Algorithm}\label{overview}
Our main result is a linear time algorithm for \lcs\ with approximation factor $O(n^{\exponent})$. We obtain this result through an array of combinatorial algorithms. In this section, we explain the intuition behind each algorithm and defer the details and proofs to Sections~\ref{bss}, \ref{bfa}, and \ref{bbrs}. Our analysis involves nontrivial applications of Dilworth's theorem, Mirsky's theorem, and  Tur\'an's theorem which may be of independent interest. Before we begin, let us first discuss how a linear time algorithm can approximate \lcs\ within a factor $\sqrt{n}$.

For a given pair of strings $s$ and $\bar{s}$, we construct a string $s^*$ from $s$ by removing each character of $s$ with probability $1-1/\sqrt{n}$. Due to the construction of $s^*$, both $|s^*| = O(\sqrt{n})$ and $|\mathsf{lcs}(s^*,\bar{s})| = \Omega(|\flcs|/\sqrt{n})$ hold w.h.p. Hence, it suffices to compute the \lcs\ of $s^*$ and $\bar{s}$. To this end, we slightly modify the conventional dynamic program for computing \lcs\ and construct a two-dimensional array $T^*$ that stores the following information

$$
T^*[i][j]= \begin{cases}
\text{ the smallest $k$ s.t.} & \hspace{-2mm}\text{ if }|\mathsf{lcs}(s^*,\bar{s})| \geq j \\
	 |\mathsf{lcs}(s^*[1,i],\bar{s}[1,k])| \geq j & \\
\infty & \hspace{-2mm}\text{ otherwise } \end{cases}
$$
Using the above definition, we can construct table $T^*$ via the following recursive formula:

\begin{equation}\label{dp2}
T^*[i][j] := \min\Big\{\begin{array}{c} T^*[i-1][j],\\ \first(\bar{s},T^*[i-1][j-1], s^*_i)\end{array}\Big\}
\end{equation} 
where $\first(\bar{s},T^*[i-1][j-1], s^*_i)$ is the index of the first occurrence of $s^*_i$ in $\bar{s}$ after position $T^*[i-1][j-1]$ (or $\infty$ if $s_i^*$ does not appear in $\bar{s}$ after position $T^*[i-1][j-1]$). Notice that both $i$ and $j$ lie in range $[0,|s^*|]$ which is bounded by $O(\sqrt{n})$.
In addition to this, $\first(\bar{s},T^*[i-1][j-1], s^*_i)$ can be computed in time $O(\log n)$ via a simple data structure and thus Algorithm \ref{alg:5} can be implemented in time $O(\sqrt{n}^2 \log n) = O(n \log n)$. 

\setcounter{algocf}{-1}
\begin{algorithm2e}
	\KwData{
		$s$ and $\bar{s}$.}
	\KwResult{A $\sqrt n$ approximate solution for \lcs.}
	$s^* \leftarrow $ an empty string\;
	\For{$i \in [1,n]$}{
		$p \leftarrow$ a random variable which is equal to $1$ with probability $1/\sqrt{n}$ and 0 otherwise \;
		\If {$p = 1$}{
			add $s_i$ to the end of $s^*$\;
		}
	}
	$T^* \leftarrow \text{\small a $[0, \ldots, |s^*|] \times [0, \ldots, |s^*|]$ array}$
	 \text{\qquad \qquad \qquad initially containing $\infty$}\;
	\For{$i \leftarrow 0$ to $|s^*|$}{
		$T^*[i][0] \leftarrow 0$\;
	}
	
	\For{$j \leftarrow 1$ to $|s^*|$}{
		\For {$i \leftarrow 1$ to $|s^*|$}{
			$T^*[i][j] \leftarrow \min\{T^*[i-1][j], \first(\bar{s},T^*[i-1][j-1], s^*_i)\}$\;
		}
	}
	
	$mx \leftarrow |s^*|$\;
	\While{$T^*[|s^*|][mx] = \infty$}{
		$mx \leftarrow mx -1$;
	}
	\Return($mx$)\;
	\caption{\textsf{a $\sqrt n$ approximate solution for \lcs.}}
	\label{alg:5}
\end{algorithm2e}


\begin{observation}
	With high probability, Algorithm \ref{alg:5} approximates $\lcs$ within a factor of $\sqrt{n}$ in near linear time $(O(n \log n))$.
\end{observation}

We improve Algorithm \ref{alg:5} through an array of combinatorial algorithms. To keep the analysis simple, for now, we only focus on beating the $\sqrt{n}$ barrier and obtaining an approximation factor which is better than $\sqrt{n}$ by a multiplicative factor of $n^{\epsilon}$ for some $\epsilon > 0$. Thus, for the sake of simplicity, we define the notation $\bar{O}$ which is similar to the $O$ notation except that $\bar{O}$ hides all $n^{o(1)}$ factors.

\begin{definition}
	$$f(n) \in \tildbar(g(n))  \iff  f(n) \in O(n^{o(1)}g(n)).$$
\end{definition}

We similarly define notations $\bar{o}, \bar{\Omega},$ and $\bar{\omega}$ as follows: 

\begin{definition}
	$$f(n) \in \bar{o}(g(n))  \iff  g(n) \notin \tildbar(f(n))$$\\\vspace{-1cm}
	$$f(n) \in \bar{\Omega}(g(n))  \iff  g(n) \in \tildbar(f(n))$$\\\vspace{-1cm}
	$$f(n) \in \bar{\omega}(g(n))  \iff  f(n) \notin \tildbar(g(n))$$\\\vspace{-1cm}
\end{definition}

In the rest of this section, we use the above definitions to demonstrate the ideas of our solution in an informal way. Based on the above definitions, we aim to design a linear time algorithm with approximation factor $\bar{o}(\sqrt{n})$. Before we go further, we note a few points.
\begin{itemize}
	\item We emphasize that the arguments we make in this section are \textbf{not} mathematically coherent and we may intentionally oversimplify some issues here in the hope that the reader grasps a more intuitive explanation of the ideas. Later in Sections \ref{bss}, \ref{bfa}, and \ref{bbrs} we bring formal and detailed discussions to prove the approximation factor of our solution. 
	\item For simplicity, in the pseudocodes of the algorithms, we only report the solution size. However, it is easy to observe that in each case the actual solution can also be reported in linear time.
	\item Anywhere we use the term w.h.p. we mean a probability whose difference with 1 is exponentially small (in terms of $n$).
	\item It may be confusing how an exponent $\exponent$ is obtained using the algorithms that we outline in this section. However, a patient reader can find a lot of technical hurdles in Sections~\ref{bss}, \ref{bfa}, and \ref{bbrs} that we skip or simplify in this section. By considering these technicalities, the upper bound on the exponent of the approximation factor would be equal to $\exponent$.
\end{itemize}
 As mentioned before, our solution consists of several algorithms that we run one by one. Each of the algorithms performs well on a class of inputs. At each step of our solution, we use the term \textit{nontrivial} instances to refer to the inputs that are not approximated within a factor of $\bar{o}(\sqrt{n})$ by the algorithms that have been performed until that step. Therefore, initially, all the input instances are nontrivial. 

The first two algorithms that we perform on the input instances are Algorithms \ref{alg:2} and \ref{alg:7}. Algorithm \ref{alg:2} is a modification of Algorithm \ref{alg:5} which is $\bar{o}(\sqrt{n})$ approximation when the solution size is small. Here we skip the details and just mention that after applying Algorithm \ref{alg:2}, the solution size in the nontrivial instances is lower bounded by $\bar{\Omega}(n)$.
\begin{observation}
	After running Algorithm \ref{alg:2} we have $|\mathsf{lcs}(s,\bar{s})| = \bar{\Omega}(n)$ for all the nontrivial instances.
\end{observation}
Next, we discuss Algorithm \ref{alg:7} in Section \ref{bfa} and show that after running Algorithm \ref{alg:7}, the number of symbols in the nontrivial instances is bounded $\bar{O}(n^{1/2})$.
\begin{observation}
	After running Algorithm \ref{alg:7}, we have $m = \bar{O}(n^{1/2})$ for any nontrivial instance.
\end{observation}

The more technically involved part of the analysis is dedicated to Algorithms \ref{alg:3}, and \ref{alg:4} which we discuss in Section \ref{sec:overview:4}.

\subsection{Decomposition into Blocks \& Permutations}
\label{sec:overview:4}
The last step of our algorithm is technically more involved as we perform two separate algorithms and argue that the better of the two algorithms has an approximation factor of $\bar{o}(\sqrt{n})$ on the nontrivial instances. So far, we have shown that in the nontrivial instances we have $m = \bar{O}(\sqrt{n})$ and $|\mathsf{lcs}(s,\bar{s})| = \bar{\Omega}(n)$. Therefore, we assume that the rest of the cases are excluded from the nontrivial instances.

We divide both $s$ and $\bar{s}$ into blocks of size $\sqrt{n}$. Denote by $\block_i$ and $\barblock_i$, the $i$'th block of $s$ and $\bar{s}$, respectively. We perform Algorithms \ref{alg:3} and \ref{alg:4} simultaneously on $s$ and $\bar{s}$ and argue that one of them performs well on each nontrivial instance. We first bring the pseudocodes of Algorithms \ref{alg:3} and \ref{alg:4} and then present the intuition behind each algorithm.
\setcounter{algocf}{2}
\input{btb}

\input{rash}

As mentioned earlier, we decompose the strings into $\sqrt{n}$ blocks of equal size in both the algorithms. In Algorithm \ref{alg:3}, we estimate the \lcs\ of each pair of blocks from $s$ and $\bar{s}$ by a very simple sampling argument and then run a dynamic program to construct a solution that incorporates all blocks. In Algorithm \ref{alg:4} however, we first turn each block into a \textit{semi-permutation}. That is, we remove some characters from each block to make sure no symbol appears more than once. Next, we choose a random integer $r$ uniformly from $[\sqrt{n}]$. We then construct a solution in which each block $\block_i$ is associated to block $\barblock_{i+r}$. Since the blocks are all semi-permutations, we can compute their \lcs's in linear time and therefore we can run Algorithm \ref{alg:4} in near linear time. The details about the running times of Algorithms \ref{alg:3} and \ref{alg:4} can be found in Section \ref{bbrs}.

\input{windows6} 

To better understand the relation between Algorithms \ref{alg:3} and \ref{alg:4}, fix an optimal solution $\opt$ to be the longest common subsequence for a nontrivial instance of $s$ and $\bar{s}$. We can break $\opt$ into at most $2\sqrt{n}$ parts such that the corresponding characters of each part lie within a single block in both of the strings. If a symbol $c$ appears at least $\bar{\omega}(1)$ times in one part of $\opt$, then we call the corresponding characters of that part \textit{crebris}. We call the rest of the characters of $\opt$ \textit{rarus}
\footnote{Crebris and rarus are Latin equivalents of words frequent and infrequent}
. Algorithm \ref{alg:3} aims to approximate the crebris characters of the optimal solution and Algorithm \ref{alg:4} focuses on the rarus characters. Since either the crebris or rarus pieces of the optimal solution provide a $1/2$ guarantee, any $\bar{o}(\sqrt{n})$ solution for each set of characters gives us an $\bar{o}(\sqrt{n})$ solution to \lcs. 

We argue in Section \ref{bbrs} that Algorithm \ref{alg:3} provides an $\bar{o}(\sqrt{n})$ approximation of the crebris characters. In other words, we show that the length of the solution that Algorithm \ref{alg:3} finds is at least $\bar{\omega}(\sqrt{n})$, if the crebris characters dominate the solution. Let $c$ be a crebris character  in a part of the solution which is associated to blocks $\block_i$ and $\barblock_j$. Since $c$ is crebris, then $\bar{\omega}(1)$ characters of these blocks have the same symbol as $c$ and therefore the expected value of $T[i][j]$ in Algorithm \ref{alg:3} is an $\bar{o}(\sqrt{n})$ approximation of $|\mathsf{lcs}(\block_i,\barblock_j)|$. This observation enables us to prove that Algorithm \ref{alg:3} gives an $\bar{o}(\sqrt{n})$ approximation of the crebris characters of $\opt$.

The analysis for Algorithm \ref{alg:4} is much more technically involved. Notice that after applying Algorithm \ref{alg:3}, we know that in the nontrivial instances, for any optimal solution $\opt$, the majority of the characters are rarus. The nice property of the rarus characters is that after modifying each block into a semi-permutation, the size of the rarus characters in any optimal solution remains relatively intact. Notice that some rarus characters may appear too many ($\bar{\omega}(1)$) times in both corresponding blocks of $s$ and $\bar{s}$, however, the majority of the rarus characters appear only a few times $(\bar{O}(1))$ in the corresponding blocks and therefore after turning blocks into semi-permutations, the size of the rarus portion of any optimal solution remains relatively unchanged.

Let us refer to the semi-permutations by refined blocks and denote them by $\block'_i$ and $\barblock'_i$. We also denote the refined strings by $s'$ and $\bar{s'}$. We show in Section \ref{bbrs} that if $m = \bar{O}(\sqrt n)$ and $|\mathsf{lcs}(s',\bar{s'})| = \bar{\Omega}(n)$ then we have 
$$\mathbb{E}_{i,j \sim [\sqrt{n}]}[\mathsf{lcs}(\block'_i,\barblock'_j)] = \bar{\omega}(1)$$
for two randomly selected refined blocks $\block_i$ and $\block'_j$. This indeed proves that Algorithm \ref{alg:4} obtains an $\bar{o}(\sqrt{n})$ approximation of the optimal solution, provided that all the previous algorithms have failed to do so.

Our analysis is based on two well-known combinatorial theorems: Dilworth's theorem and Tur\'an's theorem.

\vspace{0.2cm}
{\noindent T{\footnotesize HEOREM} (Tur\'an 1941 \cite{turan1941external}). \textit{  Let $G$ be any graph with $n$ vertices, such that $G$ is $K_{r+1}$-free. Then the number of edges in G is at most $(1-1/r)n^2/2.$}
\vspace{0.2cm}

{\noindent T{\footnotesize HEOREM} (Dilworth-1950 \cite{dilworth1950decomposition}). \textit{  	Let $P$ be a poset. Then there exists an antichain $X$ and a chain decomposition $Y$ satisfying $|X| = |Y|$.}}
\vspace{0.2cm}

For a formal definition of poset, antichain, and chain decomposition, we refer the reader to Section \ref{bbrs}. 

To illustrate the idea, let us consider a simple case in which $|\Sigma|=\sqrt n$ and each block is a complete permutation of $\Sigma$. Also, we assume for simplicity that in the optimal solution, every block $b_i$ of $s$ is matched with block $b'_i$ of $\sbar$, that is $|\mathsf{lcs}(\block_i, \barblock_i)|= \bar{\Omega}(\sqrt n)$ for each $1 \leq i \leq \sqrt{n}$. We show that in this case, the expected size of \lcs\ for two randomly selected blocks of $s'$ and $\bar{s'}$ is $\bar{\Omega}(n^{1/6})$. In this case, we first leverage  Dilworth's theorem to prove Lemma \ref{cor:1}.

\vspace{0.2cm}
{\noindent \textbf{Lemma} \ref{cor:1} (restated). \textit{Let $\pi_1$, $\pi_2$, and $\pi_3$ be three permutations of numbers $[1 \ldots m]$. Then, we have $$\max\{|\mathsf{lcs}(\pi_1, \pi_2)|, |\mathsf{lcs}(\pi_1, \pi_3)|, |\mathsf{lcs}(\pi_2, \pi_3)|\} \geq m^{1/3}.$$}

Now, construct a graph $G$ in which we have a vertex for every $1 \leq i \leq \sqrt{n}$ and there is an edge between two vertices $i$ and $j$ if and only if $|\mathsf{lcs}(\block_i, \barblock_j)| + |\mathsf{lcs}(\barblock_i, \block_j)| \geq \bar{\Omega}(n^{1/6})$. Using Lemma \ref{cor:1} we show that for any three vertices $i$, $j$, and $k$, at least there is an edge between one pair. Next, we apply the Tur\'an's theorem to show that $G$ is dense and thus the expected size of the \lcs\ for two randomly selected blocks of $s'$ and $\bar{s'}$ is at least $\bar{\Omega}(n^{1/6})$.  This implies that the expected size of the solution returned by Algorithm \ref{alg:4} is $\bar{\Omega}(n^{1/6+1/2}) = \bar{\Omega}(n^{2/3})$ which is $\bar{O}(n^{1/3})$ approximation since the size of the optimal solution is bounded by $n$.

Indeed the proof for the case of complete permutation is much cleaner than the case of semi-permutation. However, we present a nice reduction in Section \ref{bbrs} that shows the argument carries over to the semi-permutation case. Roughly speaking, the idea is to create a \textit{mask} for both refined strings $s'$ and $\bar{s}'$ and prove some nice properties for the masks. Each mask is essentially a permutation of the symbols in $\Sigma$ chosen in a way that the size of the \lcs\ of the two masks is tentatively equal to the size of the  \lcs\ of two random blocks chosen from $s'$ and $\bar{s'}$. We then use the masks to turn each semi-permutation into a complete permutation by padding the characters that do not appear in the block to the end of it, according to their order in the corresponding mask. After this, each block is a complete permutation and the lemma holds for it. We then argue that the lemma also holds for the original refined blocks.

Let us be more precise about the proof. In Algorithm \ref{alg:4}, after refining the blocks  into semi-permutations, we know that in the nontrivial instances, the size of the solution is at least $\bar{\Omega}(n)$. Now, for a symbol $c \in \Sigma$, define $p_c$ to be the probability that a random block (of any string) contains symbol $c$. If $p_c = \bar{o}(1)$ for some character, then it is safe to assume that symbol $c$ does not contribute significantly to the solution. Thus, after removing all such symbols, the solution size is still $\bar{\Omega}(n)$. Notice that we do not remove the characters in our algorithm and just for the case of simplicity, we ignore these characters in our analysis. 

Now, for every symbol $c \in \Sigma$ we have the property that $p_c = \bar{\Omega}(1)$. Thus, if we take $\bar{O}(1)$ (but large enough)  random blocks of the strings, each symbol appears in at least one block w.h.p. Based on this idea, we construct a mask for each string as follows: we start with an empty mask. Each time we randomly select a block of that string and append the characters that are not already in the mask but appear in the selected block to the end of the mask. The order of the newly added characters is the same as their order in the selected block. We repeat this procedure $\bar{O}(1)$ times to make sure the mask contains all of the symbols. 

\input{mask}

Again we emphasize that the construction of the masks is only part of the proof and does not play any role in Algorithm \ref{alg:4}. Denote the constructed mask for string $s$ by $\mask{s}$ and the one for string $\bar{s}$ by $\mask{\bar{s}}$. If $\mathbb{E}[|\mathsf{lcs}(\mask{s},\mask{\bar{s}})|] = \bar{\omega}(1)$ then we can imply that for two randomly selected blocks $b'_i$ of $s$ and $\bar{b'}_j$ of $\bar{s}$ we have  $\mathbb{E}[|\mathsf{lcs}(b'_i,\bar{b'_j})|] = \bar{\omega}(1)$ and thus the expected solution size of Algorithm \ref{alg:4} is $\bar{\omega}(\sqrt{n})$. This implies that Algorithm \ref{alg:4} is $\bar{o}(\sqrt{n})$ approximation. Therefore, we assume w.l.o.g. that $\mathbb{E}[|\mathsf{lcs}(b'_i,\bar{b'_j})|] = \bar{O}(1)$.

Now, for every refined block $b'_i$ of $s$ we construct a complete permutation by padding the characters that are not included in $b'_i$ to the end of it. However, we maintain the property that the order of the newly added characters should be exactly the same as their order in $\mask{s}$. Let us refer to these (complete) permutations by $\cp_i$ for $s$. We similarly transform the semi-permutations of $\bar{s}$ into complete permutations using $\mask{\bar{s}}$. We denote the complete permutations for $\bar{s}$ by $\bar{\cp}_i$. Notice that again, the construction of the complete permutations is only for the sake of analysis and we do not explicitly construct these permutations in our algorithm.

\input{mask2}

Now, we can use the previous analysis and prove that if the size of the solution is large ($\bar{\Omega}(n)$) then the \lcs\ of two randomly selected complete permutations is large ($\bar{\omega}(1))$. However, this does not give us a bound for the refined blocks since complete permutations have extra characters and those characters may contribute to a large portion of  the common subsequence. Here, we leverage the fact that $\mathbb{E}[|\mathsf{lcs}(\mask{s},\mask{\sbar})|] = \bar{O}(1)$ or in other words, the expected size of the longest common subsequence for the masks is small. Using this bound, we can show that the contribution of the extra characters to the common subsequence is bounded by $\bar{\omega}(1)$ and therefore the majority of the solution is due to the characters of the refined blocks. This proves that the \lcs\ of two randomly selected refined blocks of $s$ and $\bar{s}$ is large ($\bar{\omega}(1))$ and thus Algorithm \ref{alg:4} is $\bar{o}(\sqrt{n})$ approximate.

%% file: btb.tex
\begin{algorithm2e}[h!]
	\KwData{
		$s$ and $\bar{s}$.}
	\KwResult{An approximate solution for \lcs.}
	Divide $s$ and $\bar{s}$ into $\sqrt{n}$ blocks as explained;
	\Comment{$b_i$: $i$'th block of $s$ and $\barblock_j$: $j$'th block of $\sbar$}\\
	Compute $\fr_{c}(\block_i)$ and $\fr_c(\barblock_i)$ for every symbol $c \in \Sigma$ and $i \in [\sqrt{n}]$\;
	$T,D \leftarrow \text{ two } \sqrt{n} \times \sqrt{n}$ tables\;
	\For{$i \leftarrow 1$ to $\sqrt n$}{
		\For{$j\leftarrow 1$ to $\sqrt n$}{
			$c \leftarrow $ a random character of block $\block_i$\;
			$T[i][j] \leftarrow \min\{ \fr_c(\block_i), \fr_c(\barblock_j) \}$\;
		}
	}
	\For{$i \leftarrow 1$ to $\sqrt n$}{
		\For{$j \leftarrow 1$ to $\sqrt n$}{
			$D[i][j] \leftarrow \max\{ D[i][j-1], D[i-1][j], D[i-1][j-1] + T[i][j] \}$\;
		}
	}
	\textbf{Return}($\max_{i,j} D[i][j]$)
	\caption{\blocktobloc}
	\label{alg:3}
\end{algorithm2e}

%% file: rash.tex
\begin{algorithm2e}[h!]
	\KwData{
		$s$ and $\bar{s}$}
	\KwResult{An approximate solution for \lcs.}
	Divide $s$ and $\bar{s}$ into $\sqrt{n}$ blocks as explained\;
	\For{$i \in [\sqrt{n}]$}{
		\For{each symbol $c$ that appears in $\block_i$}{
			randomly choose one appearance and remove the rest of the appearances of $c$ from $\block_i$\;
		}
		\For{each symbol $c$ that appears in $\barblock_i$}{
			randomly choose one appearance and remove the rest of the appearances of $c$ from $\barblock_i$\;
		}
	}
	$r \leftarrow $a random integer number between $1$ and $\sqrt{n}$.\;
	$A,B \leftarrow 0$\;
	\For{$i = 1$ to  $\sqrt n - r$ }{
		$A \leftarrow A + |\llcs(\block_i,\barblock_{i+r})|$\;
	}
	\For{$i = \sqrt{n}-r+1$ to  $\sqrt n$ }{
		$B \leftarrow B + |\llcs(\block_i,\barblock_{i+r-\sqrt{n}})|$\;
	}
	\Return $\max\{A,B\}$
	\caption{\rash}
	\label{alg:4}
\end{algorithm2e}

%% file: windows6.tex
\begin{figure*}[h!]

\begin{center}
\begin{tikzpicture}[scale=0.86, transform shape]

\newcommand\Yonedown{0}
\newcommand\Ymiddledown{-1.8}
\newcommand\Ytwodown{-3.5}

\draw (0,\Yonedown) -- (17.5,\Yonedown) -- (17.5,\Yonedown+1) -- (0,\Yonedown+1) -- (0,\Yonedown);

\draw [dashed] (0.15,\Yonedown + 0.1) rectangle (2.35,\Yonedown+0.9);
\draw [fill=gray, opacity=0.2] (0.15,\Yonedown+0.1) rectangle (2.35,\Yonedown+0.9);
\draw (2.5,\Yonedown) -- (2.5,\Yonedown+1);
\draw [dashed] (2.65,\Yonedown+0.1) rectangle (4.85,\Yonedown+0.9);
\draw [fill=gray, opacity=0.2] (2.65,\Yonedown+0.1) rectangle (4.85,\Yonedown+0.9);

\draw [dashed] (5.15,\Yonedown+0.1) rectangle (7.35,\Yonedown+0.9);
\draw [fill=gray, opacity=0.2] (5.15,\Yonedown+0.1) rectangle (7.35,\Yonedown+0.9);
\draw (7.5,\Yonedown) -- (7.5,\Yonedown+1);
\draw [dashed] (7.65,\Yonedown+0.1) rectangle (9.85,\Yonedown+0.9);
\draw [fill=gray, opacity=0.2] (7.65,\Yonedown+0.1) rectangle (9.85,\Yonedown+0.9);

\draw (15,\Yonedown) -- (15,\Yonedown+1);
\draw (5,\Yonedown) -- (5,\Yonedown+1);
\draw (10,\Yonedown) -- (10,\Yonedown+1);

\draw [fill=gray, opacity=0.2] (10.15,\Yonedown+0.1) rectangle (12.35,\Yonedown+0.9);
\draw [dashed] (10.15,\Yonedown+0.1) rectangle (12.35,\Yonedown+0.9);
\draw (12.5,\Yonedown) -- (12.5,\Yonedown+1);
\draw [fill=gray, opacity=0.2] (12.65,\Yonedown+0.1) rectangle (14.85,\Yonedown+0.9);
\draw [dashed] (12.65,\Yonedown+0.1) rectangle (14.85,\Yonedown+0.9);

\draw (15,\Yonedown) -- (15,\Yonedown+1);
\draw [fill=gray, opacity=0.2] (15.15,\Yonedown+0.1) rectangle (17.35,\Yonedown+0.9);
\draw [dashed] (15.15,\Yonedown+0.1) rectangle (17.35,\Yonedown+0.9);


\node[text width=0.3cm] at (0.4,\Yonedown+0.5) 
{   \huge $c\color{BrickRed}a\color{black}c\color{BrickRed}bbb\color{black}$ };

\node[text width=0.3cm] at (2.9,\Yonedown+0.5) 
{   \huge $ccc\color{Cyan}a\color{black}cc$ };

\node[text width=0.3cm] at (5.4,\Yonedown+0.5) 
{   \huge $cccc\color{Cyan}b\color{black}c$ };

\node[text width=0.3cm] at (7.9,\Yonedown+0.5) 
{   \huge $cc\color{Cyan}b\color{black}c\color{BrickRed}ab\color{black}$ };

\node[text width=0.3cm] at (10.4,\Yonedown+0.5) 
{   \huge $\color{BrickRed}b\color{black}cccc$ };

\node[text width=0.3cm] at (12.9,\Yonedown+0.5) 
{   \huge $ccc\color{BrickRed}a\color{black}cc$ };

\node[text width=0.3cm] at (15.4,\Yonedown+0.5) 
{   \huge $\color{Cyan}a\color{black}ccc\color{Cyan}aa$ };


\node[text width=3.3cm] at (8,\Ymiddledown+0.5) 
{   \huge $\color{black}|\color{BrickRed}abbb\color{black}|\color{Cyan}abb\color{black}|\color{BrickRed}abba\color{black}|\color{Cyan}aaa\color{black}|$ };



\node[text width=0.3cm] at (0.4,\Ytwodown+0.5) 
{   \huge $d\color{BrickRed}a\color{black}dd\color{BrickRed}b\color{black}$ };

\node[text width=0.3cm] at (2.9,\Ytwodown+0.5) 
{   \huge $ddd\color{BrickRed}b\color{black}d$ };

\node[text width=0.3cm] at (5.4,\Ytwodown+0.5) 
{   \huge $d\color{BrickRed}b\color{Cyan}abb\color{black}$ };

\node[text width=0.3cm] at (7.9,\Ytwodown+0.5) 
{   \huge $\color{BrickRed}abba\color{black}dd$ };

\node[text width=0.3cm] at (10.4,\Ytwodown+0.5) 
{   \huge $ddddd$ };

\node[text width=0.3cm] at (12.9,\Ytwodown+0.5) 
{   \huge $dd\color{Cyan}a\color{black}dd$ };

\node[text width=0.3cm] at (15.4,\Ytwodown+0.5) 
{   \huge $\color{Cyan}aa\color{black}ddd$ };

\draw (0,\Ytwodown) -- (17.5,\Ytwodown) -- (17.5,\Ytwodown+1) -- (0,\Ytwodown+1) -- (0,\Ytwodown);

\draw [dashed] (0.15,\Ytwodown + 0.1) rectangle (2.35,\Ytwodown+0.9);
\draw [fill=gray, opacity=0.2] (0.15,\Ytwodown+0.1) rectangle (2.35,\Ytwodown+0.9);
\draw (2.5,\Ytwodown) -- (2.5,\Ytwodown+1);
\draw [dashed] (2.65,\Ytwodown+0.1) rectangle (4.85,\Ytwodown+0.9);
\draw [fill=gray, opacity=0.2] (2.65,\Ytwodown+0.1) rectangle (4.85,\Ytwodown+0.9);

\draw [dashed] (5.15,\Ytwodown+0.1) rectangle (7.35,\Ytwodown+0.9);
\draw [fill=gray, opacity=0.2] (5.15,\Ytwodown+0.1) rectangle (7.35,\Ytwodown+0.9);
\draw (7.5,\Ytwodown) -- (7.5,\Ytwodown+1);
\draw [dashed] (7.65,\Ytwodown+0.1) rectangle (9.85,\Ytwodown+0.9);
\draw [fill=gray, opacity=0.2] (7.65,\Ytwodown+0.1) rectangle (9.85,\Ytwodown+0.9);

\draw (15,\Ytwodown) -- (15,\Ytwodown+1);
\draw (5,\Ytwodown) -- (5,\Ytwodown+1);
\draw (10,\Ytwodown) -- (10,\Ytwodown+1);

\draw [fill=gray, opacity=0.2] (10.15,\Ytwodown+0.1) rectangle (12.35,\Ytwodown+0.9);
\draw [dashed] (10.15,\Ytwodown+0.1) rectangle (12.35,\Ytwodown+0.9);
\draw (12.5,\Ytwodown) -- (12.5,\Ytwodown+1);
\draw [fill=gray, opacity=0.2] (12.65,\Ytwodown+0.1) rectangle (14.85,\Ytwodown+0.9);
\draw [dashed] (12.65,\Ytwodown+0.1) rectangle (14.85,\Ytwodown+0.9);

\draw (15,\Ytwodown) -- (15,\Ytwodown+1);
\draw [fill=gray, opacity=0.2] (15.15,\Ytwodown+0.1) rectangle (17.35,\Ytwodown+0.9);
\draw [dashed] (15.15,\Ytwodown+0.1) rectangle (17.35,\Ytwodown+0.9);

\node[text width=0.5cm] at (-0.40,\Yonedown + 0.4) 
{   \huge $s$ };

\node[text width=0.5cm] at (-0.40,\Ytwodown + 0.4) 
{   \huge $\bar{s}$ };

\node[text width=0.5cm] at (5.1,\Ymiddledown + 0.5) 
{   \huge $\opt$: };

\end{tikzpicture}
\end{center}
\caption{An example of the decomposition of the optimal solution into several parts.}\label{fig:windows6}
\end{figure*}
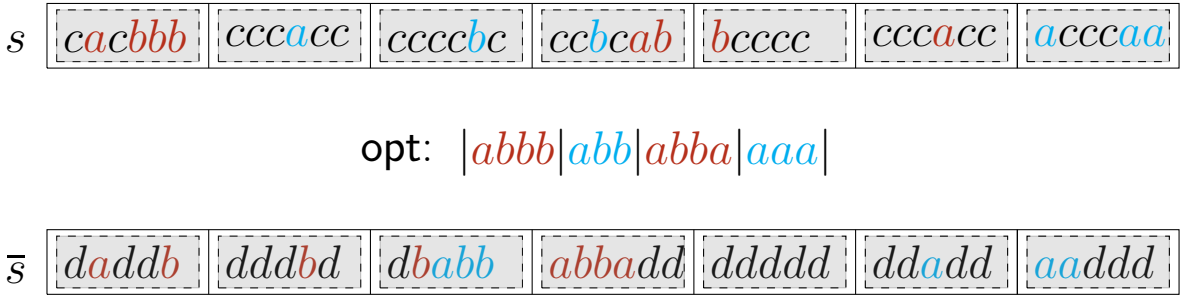

%% file: mask.tex
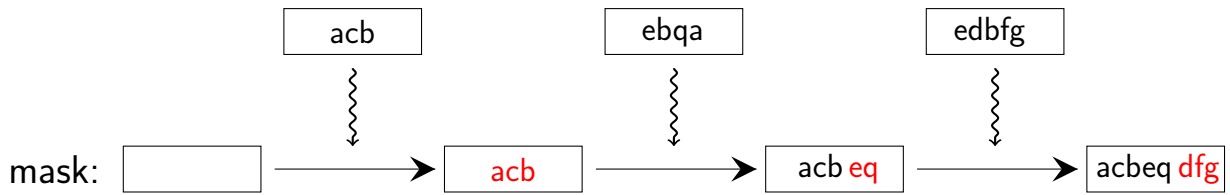
\begin{figure*}[h!]

\begin{center}
\hypersetup{linkcolor=red}
	
\begin{tikzpicture}[scale=0.61, transform shape]
\definecolor{mygray}{gray}{0.95}
\definecolor{myblue}{rgb}{1,0.9,0.9}

\tikzstyle{mybox} = [draw=black, fill=mygray, very thick,
rectangle, rounded corners, inner sep=10pt, inner ysep=20pt]
\tikzstyle{fancytitle} =[fill=black, text=white, ellipse]
\tikzset{myptr/.style={decoration={markings,mark=at position 1 with {\arrow[scale=3,>=stealth]{>}}},postaction={decorate}}}
\tikzset{myptr2/.style={decoration={markings,mark=at position 1 with {\arrow[scale=2,>=stealth]{>}}},postaction={decorate}}}

\draw (0,0) rectangle (3,1);
\draw [myptr] (3.3,0.5) -- (6.8,0.5);
\draw [thick,->,decorate,decoration={snake,amplitude=.4mm,segment length=2mm,post length=1mm}]
(5,2.7) -- (5,1);
\draw (3.5,3) rectangle (6.5,4);

\draw (7,0) rectangle (10,1);
\draw [myptr] (10.3,0.5) -- (13.8,0.5);
\draw [thick,->,decorate,decoration={snake,amplitude=.4mm,segment length=2mm,post length=1mm}]
 (12,2.7) -- (12,1);
\draw (10.5,3) rectangle (13.5,4);

\draw (14,0) rectangle (17,1);
\draw [myptr] (17.3,0.5) -- (20.8,0.5);
\draw [thick,->,decorate,decoration={snake,amplitude=.4mm,segment length=2mm,post length=1mm}]
(19,2.7) -- (19,1);
\draw (17.5,3) rectangle (20.5,4);

\draw (21,0) rectangle (24,1);

\node[text width=3cm] at (2,0.5) {    };
\node[text width=3cm] at (9.5,0.5) { \huge \color{red} \textsf{acb} \color{black} };
\node[text width=3cm] at (6,3.5) { \huge  \textsf{acb}  };
\node[text width=3cm] at (12.8,3.5) { \huge  \textsf{ebqa}  };
\node[text width=3cm] at (16.2,0.5) { \huge  \textsf{acb} \color{red} \textsf{eq}\color{black} };
\node[text width=3cm] at (19.6,3.5) { \huge  \textsf{edbfg}  };
\node[text width=3cm] at (22.7,0.5) { \huge  \textsf{acbeq} \color{red} \textsf{dfg}\color{black} };

\node[text width=3cm] at (-1,0.5) { \Huge  \textsf{mask:} };
\end{tikzpicture}
\end{center}
\caption{An example for the construction of the masks}\label{fig:mask}
\end{figure*}
\hypersetup{linkcolor=red}

%% file: mask2.tex
\begin{figure*}[h!]

\begin{center}
\hypersetup{linkcolor=red}
	
\begin{tikzpicture}[scale=0.61, transform shape]
\definecolor{mygray}{gray}{0.95}
\definecolor{myblue}{rgb}{1,0.9,0.9}

\tikzstyle{mybox} = [draw=black, fill=mygray, very thick,
rectangle, rounded corners, inner sep=10pt, inner ysep=20pt]
\tikzstyle{fancytitle} =[fill=black, text=white, ellipse]
\tikzset{myptr/.style={decoration={markings,mark=at position 1 with {\arrow[scale=3,>=stealth]{>}}},postaction={decorate}}}
\tikzset{myptr2/.style={decoration={markings,mark=at position 1 with {\arrow[scale=2,>=stealth]{>}}},postaction={decorate}}}

\draw (3.5,3) rectangle (6.5,4);
\draw (10.5,3) rectangle (13.5,4);
\draw (17.5,3) rectangle (20.5,4);

\draw[myptr2] (5,2.8) -- (11.8,1.2);
\draw[myptr2] (12,2.8) -- (12,1.2);
\draw[myptr2] (19,2.8) -- (12.2,1.2);

\draw (3.5,-3) rectangle (6.5,-2);
\draw (10.5,-3) rectangle (13.5,-2);
\draw (17.5,-3) rectangle (20.5,-2);

\draw [myptr] (12,-0.2) -- (5,-1.8); 
\draw [myptr] (12,-0.2) -- (12,-1.8); 
\draw [myptr] (12,-0.2) -- (19,-1.8); 

\node[text width=3cm] at (6,3.5) { \huge  \textsf{acb}  };
\node[text width=3cm] at (12.8,3.5) { \huge  \textsf{ebqa}  };
\node[text width=3cm] at (19.6,3.5) { \huge  \textsf{edbfg}  };

\node[text width=3cm] at (5.2,-2.5) { \huge  \textsf{acb} \color{red}\textsf{eqdfg}\color{black} };
\node[text width=3cm] at (12.2,-2.5) { \huge  \textsf{ebqa} \color{red}\textsf{cdfg}\color{black}  };
\node[text width=3cm] at (19.2,-2.5) { \huge  \textsf{edbfg}\color{red}\textsf{acq}\color{black}  };

\node[text width=3cm] at (9.3,0.5) { \Huge  \textsf{mask:} };
\draw (10.5,0) rectangle (13.5,1);
\node[text width=3cm] at (12.2,0.5) { \huge  \textsf{acbeqdfg}  };

\end{tikzpicture}
\end{center}
\caption{An example for the construction of the masks}\label{fig:mask}
\end{figure*}
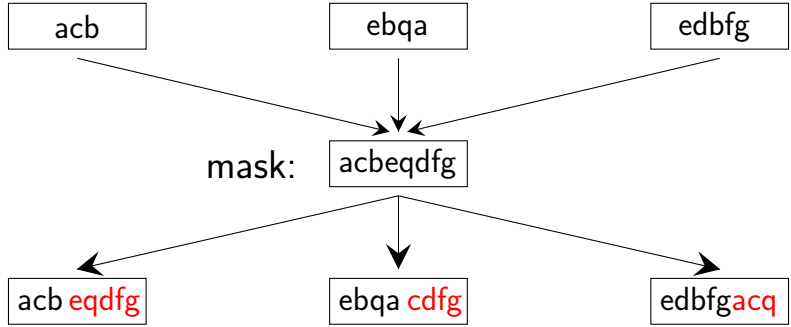
\hypersetup{linkcolor=red}

%% file: formal.tex
\section{A Linear Time $O(n^{\exponent})$ Approximation  Algorithm for \lcs} \label{sec:overview:5}
In this section we bring a formal proof to show that our algorithm is $O(n^{\exponent})$ approximation. In Sections \ref{bss}, \ref{bfa}, and \ref{bbrs} we analyze Algorithms \ref{alg:2}, \ref{alg:7}, \ref{alg:3}, and \ref{alg:4} in details and prove Lemmas \ref{lem:bss}, \ref{lem:sec6}, and \ref{lem:alg8} (a simple statement for each lemma is given in Figure \ref{fig:algorithm}). We consider two parameters $\delta$ and $\eta$ in our analysis to denote the value of $|\mathsf{lcs}(s,\bar{s})|$ by $n^{1-\delta}$ and the number of symbols by $n^{1/2+\eta}$. Therefore, we have $0 \leq \delta \leq 1$ and $-0.5 \leq \eta \leq 0.5$. As shown in Figure \ref{fig:algorithm}, the approximation factor of our algorithm is bounded by $\tilde{O}(n^\nu)$ where:
$$ \nu = \min \Bigg\{\begin{array}{c} {1/2-\delta/2},\\ {1/2-\eta}, \\ 1/2 - 1/37 + (221/37)\delta+(10/37)\eta \end{array}\Bigg\}$$ 

 Thus, we can optimize $\nu$ by solving the following linear program:

\begin{equation}
\begin{array}{ll}
\text{maximize}  & \nu\\
\text{subject to}& \nu \leq 1/2-\eta\\
	 & \nu \leq 1/2-\delta/2 \\
	 & \nu \leq 1/2 - 1/37 + (221/37)\delta+(10/37)\eta \\
	 & 0 \leq \delta \leq 1 \\
	 & -0.5 \leq \eta \leq 0.5 \\
\end{array}
\label{lp:lp}
\end{equation}

which is maximized for $\nu \simeq \exponent$, $\delta \simeq 0.004090$, and $\eta \simeq 0.002045$. This gives us the main contribution of our work.
\begin{theorem}
	\label{main}
	There exists a linear time algorithm that approximates \lcs\ within a factor $O(n^{\exponent})$ (in expectation).
\end{theorem}
\begin{proof}
	The details are shown in Figure \ref{fig:algorithm}. Here we just mention that $\exponent$ is strictly more than the optimal solution of LP \ref{lp:lp} and therefore we can ignore all the polylogarithmic factors in the approximation factor. In addition to this, a polylogarithmic factor in the running time can be handled by sampling each character of the input by a polylogarithmic rate and reducing the input size by any polylogarithmic factor. This only hurts the solution size by a polylogarithmic factor.
\end{proof}

 
\input{alltogether}

%% file: alltogether.tex
\begin{figure*}

\begin{center}
\hypersetup{linkcolor=red}
	
\begin{tikzpicture}[scale=0.65, transform shape]
\definecolor{mygray}{gray}{0.95}
\definecolor{myblue}{rgb}{1,0.9,0.9}
\draw [dashed,color=red,fill = myblue] (3.7, 5.3) -- (9.3, 5.3) -- (9.3, 3.3) -- (19.6,3.3) -- (19.6,0.7) -- (3.7, 0.7) -- (3.7, 5.3);

\tikzstyle{mybox} = [draw=black, fill=mygray, very thick,
rectangle, rounded corners, inner sep=10pt, inner ysep=20pt]
\tikzstyle{fancytitle} =[fill=black, text=white, ellipse]
\tikzset{myptr/.style={decoration={markings,mark=at position 1 with {\arrow[scale=3,>=stealth]{>}}},postaction={decorate}}}
\tikzset{myptr2/.style={decoration={markings,mark=at position 1 with {\arrow[scale=2,>=stealth]{>}}},postaction={decorate}}}

\draw [dashed, color=red, fill = myblue] (3.7,8.3) rectangle (9.3,6.7);

\draw (18.2,10) circle (1cm);
\node[text width=0.5cm] at (18,10) 
{   \huge $s,\bar{s}$ };

\draw [myptr2, very thick] (18.10,9) -- (18.10,7.5) -- (17,7.5);
\draw [myptr2, very thick] (18.30,9) -- (18.30,4.5) -- (17,4.5);

\draw [myptr2, very thick] (18.10,1) -- (18.10,-0.5) -- (17,-0.5);
\draw [myptr2, very thick ] (18.30,1) -- (18.30,-3.5) -- (17,-3.5);

\draw [dashed, color=red, fill=myblue] (3.7,-4.3) rectangle (9.3,0.3);

\draw [fill=yellow] (12,7) rectangle (17,8);
\draw  (4,7) rectangle (9,8);

\draw [myptr2, very thick] (10.5,4.5) -- (10.5,3.8) -- (18.3,3.8) -- (18.3,3);

\draw [fill=yellow] (12,4) rectangle (17,5);
\draw  (4,4) rectangle (9,5);

\draw (18.2,2) circle (1cm);
\node[text width=0.5cm] at (17.85,2) 
{   \huge $s',\bar{s'}$ };

\node[text width=3cm] at (14.8,7.5) 
{   Algorithm \ref{alg:2} };

\node[text width=3cm] at (7.2,7.5) 
{   Output 1 };

\draw [fill = yellow](-1,-7) rectangle (1,10);
\draw (-6,-6) rectangle (-4,9);
\draw [myptr] (-1,1.5) -- (-4,1.5);
\node[text width=16cm, rotate = 90] at (0.2,4.5) 
{  \Huge {Choose\hspace{0.2cm} the\hspace{0.2cm} longest\hspace{0.2cm} solution} };

\node[text width=16cm, rotate = 90] at (-4.8,2.7) 
{  \Huge {An\hspace{0.2cm} $O(n^{\exponent})$\hspace{0.2cm} approximate\hspace{0.2cm} solution} };

\draw [myptr2, very thick] (12,7.5) -- (9,7.5);
\draw [myptr2, very thick] (4,7.5) -- (1,7.5);

\draw [myptr2, very thick] (12,4.5) -- (9,4.5);
\draw [myptr2, very thick] (4,4.5) -- (1,4.5);

\draw [myptr2, very thick] (12,-0.5) -- (9,-0.5);
\draw [myptr2, very thick] (4,-0.5) -- (1,-0.5);

\draw [myptr2, very thick] (12,-3.5) -- (9,-3.5);
\draw [myptr2, very thick] (4,-3.5) -- (1,-3.5);

\draw [color=red,dashed] (6.5,8.3) -- (6.5,11) -- (2.6,11) -- (2.6,13);

\node[text width=3cm] at (14.8,4.5) 
{   Algorithm \ref{alg:7} };

\node[text width=3cm] at (7.2,4.5) 
{   Output 2 };

\draw  (12,4) rectangle (17,5);
\draw  (4,4) rectangle (9,5);

\draw [dashed,color=red] (6.5,5.3) -- (6.5,6) -- (10,6) -- (10,11) -- (12.5,11) -- (12.5,13);

\node [mybox] (box)  at (2.5,13){%
	\begin{minipage}[t!]{0.5\textwidth}
	If $|\mathsf{lcs}(s,\bar{s})| = O(n^{1-\delta})$ then output 1 has approximation factor $O(n^{1-\delta/2})$
	\end{minipage}
};
\node[fancytitle] at (box.north) {\hspace{15mm} Lemma \ref{lem:bss} (\textsf{restated})};

\node [mybox] (box2)  at (12.5,13.5){%
	\begin{minipage}[t!]{0.5\textwidth}
	For a fixed $\eta$, either  output 2 is $O(n^{1/2-\eta})$ approximate, or any $\alpha$ approximate solution for $\mathsf{lcs}(s',\bar{s'})$ is a $4\alpha$ approxmate solution for $\mathsf{lcs}(s,\bar{s})$.\\ Moreover, $s'$ and $\bar{s'}$ have at most $n^{1/2+\eta}$ symbols.
	\end{minipage}
};
\node[fancytitle] at (box2.north) {\hspace{15mm} Lemma \ref{lem:sec6} (\textsf{restated})};

\node [mybox] (box3)  at (8.5,-8.5){%
	\begin{minipage}[t!]{0.5\textwidth}
	 If  $m = O(n^{1/2+\eta})$ and $\mathsf{lcs}(s',\bar{s'}) = \Omega(n^{1-\delta})$ then the better of outputs 3 and 4 is $$\tilde{O}(\brruntime)$$ approximate.
	\end{minipage}
};

\node[fancytitle] at (box3.north) {\hspace{15mm}Lemma \ref{lem:alg8} (\textsf{restated})};

\draw [fill=yellow] (12,-1) rectangle (17,0);
\draw  (4,-1) rectangle (9,0);

\draw [fill=yellow] (12,-4) rectangle (17,-3);
\draw  (4,-4) rectangle (9,-3);

\draw [dashed, color = red] (6.5,-4.3) -- (6.5,-5) -- (8.5,-5) -- (8.5,-7);

\node[text width=3cm] at (14.8,-0.5) 
{   Algorithm \ref{alg:3} };

\node[text width=3cm] at (7.2,-0.5) 
{   Output 3 };

\node[text width=3cm] at (14.8,-3.5) 
{   Algorithm \ref{alg:4} };

\node[text width=3cm] at (7.2,-3.5) 
{   Output 4 };

\end{tikzpicture}
\end{center}
\caption{Flowchart of the $O(n^{\exponent})$-approximation Algorithm}\label{fig:algorithm}
\end{figure*}

\hypersetup{linkcolor=red}

%% file: BoundingSolutionSize.tex
\section{Algorithm \ref{alg:2}: Lower Bound on the Solution Size} \label{bss}
In the first step, we show that the instances with small solution sizes can be approximated well in linear time. Consider two strings $s$ and $\bar{s}$ such that $|\flcs| \leq n^{1-\delta}$ for a constant $\delta>0$. We argue that for such instances, Algorithm \ref{alg:2} returns an $O(n^{(1-\delta)/2})$-approximate solution. Algorithm \ref{alg:2} is essentially a modified variant of Algorithm \ref{alg:5} which is tuned to work well for instances with low solution size. 
The precondition and guarantee of Algorithm \ref{alg:2} are given below.

	\begin{tcolorbox}
		\textsf{Inputs:} $s$ and $\bar{s}$.\\
		\textsf{Precondition:}
		\begin{itemize}
			\item $|\mathsf{lcs}(s,\bar{s})| \leq n^{1-\delta}$
		\end{itemize}
		\textsf{Approximation factor:} $O(n^{1/2- \delta/2})$
	\end{tcolorbox}

Let $s$ and $\sbar$ be the inputs to Algorithm \ref{alg:2}, with the assumption that $|\flcs| \leq n^{1-\delta}$. Algorithm \ref{alg:2}  constructs a string $s^*$ from $s$ by sampling each character of $s$ with probability $n^{-(1-\delta)/2}$. 
More precisely, $s^*$ is a substring of $s$, where each character of $s$ is removed from $s^*$ with probability $1- n^{-(1-\delta)/2}$. After sampling $s^*$, Algorithm \ref{alg:2} returns $\llcs(s^*,\sbar)$ as an approximation of $\flcs$. In this section, we analyze Algorithm \ref{alg:2} in details.  
 
First, let us determine the expected size of $s^*$ and  $\llcs(s^*,\sbar)$. Since each character of $s$ appears in $s^*$ with probability $n^{-(1-\delta)/2}$, and $|s| = n$, we have $$\mathbb{E}(|s^*|) = n\cdot n^{-(1-\delta)/2} = n^{(1+\delta)/2}.$$ 
Furthermore, for each character in $\flcs$, its corresponding character in $s$ is also appeared in $s^*$ with probability $n^{-(1-\delta)/2}$. Therefore, we have 
$$
\mathbb{E}(|\llcs(s^*,\sbar)|) \geq |\flcs|  n^{-(1-\delta)/2}.
$$

By the precondition of Algorithm \ref{alg:2}, we know that the solution size is at most $n^{1-\delta}$. Thus, in order to find a $n^{(1-\delta)/2}$ approximation solution for $\flcs$, it suffices to find the longest common subsequence of $s^*$ and $\sbar$ up to a size at most $$n^{1-\delta}\cdot n^{(1-\delta)/2} =  n^{(1-\delta)/2}.$$ Therefore, using a dynamic program based on  Formula \eqref{dp2}, we can find a solution in time 
\begin{align}
|s^*|  n^{(1-\delta)/2} \log n &= n^{(1+\delta)/2}  n^{(1-\delta)/2}  \log n \nonumber\\ &= \tilde{O}(n). \label{ptime} 
\end{align}

\setcounter{algocf}{0}
\begin{algorithm2e}
	\KwData{
		$s$ and $\bar{s}$.}
	\KwResult{An approximate solution for \lcs.}
	$s^* \leftarrow $ an empty string\;
	\For{$i \in [1,n]$}{
		$p \leftarrow$ a random variable which is equal to $1$ with probability $n^{-(1-\delta)/2}$ and 0 otherwise\;
		\If {$p = 1$}{
			add $s_i$ to the end of $s^*$\;
		}
	}
	\For{$j \leftarrow 1$ to $|s^*|$}{
		$T^*[0][j] \leftarrow \infty$\;
	}
	$T^*[0][0] \leftarrow 0$\;
	\For{$j \leftarrow 1$ to $|s^*|$}{
		\For {$i \leftarrow 1$ to $n^{(1-\delta)/2}$}{
			$T^*[i][j] \leftarrow \min\{T^*[i-1][j], \first(\bar{s},T^*[i-1][j-1], s^*_i)\}$\;
		}
	}
	$mx \leftarrow n^{(1-\delta)/2}$\;
	\While{$T^*[|s^*|][mx] = \infty$}{
		$mx \leftarrow mx -1$;
	}
	\Return($mx$)\;
	\caption{\boundingthesize}
	\label{alg:2}
\end{algorithm2e}

The $\log n$ factor in Equation \eqref{ptime} is because we need to compute  $\first(\bar{s},T^*[i-1][j-1],s^*)$ for every $i,j$. Recall that $\first(\bar{s},T^*[i-1][j-1], s^*_i)$ is the index of the first occurrence of symbol $s^*_i$ in $\bar{s}$ after position $T^*[i-1][j-1]$.  Using an $O(n)$ time preprocess on $\sbar$ we can store for every symbol $c$, a sorted list of its occurrences in  $\sbar$, which can be later used to compute $\first(\bar{s},T^*[i-1][j-1],s^*)$ via a binary search in time $O(\log n)$. This, implies Lemma \ref{lem:bss}.


\begin{lemma} 
	\label{lem:bss} 
	Assuming $|\flcs| \leq n^{1-\delta}$,  Algorithm \ref{alg:2} returns an $O(n^{(1 - \delta)/2})$ approximation solution in time $O(n\log n)$.
\end{lemma}

%% file: BoundingFrequencies.tex
\section{Algorithm \ref{alg:7}: Bounding the Number of Symbols}\label{bfa}
In this step of the algorithm, we modify the input to guarantee an upper bound on the number of symbols. Notice that unlike the previous step, in this step, we may modify the input to meet this criterion while keeping the promise that the solution remains relatively intact.

Let $R := |\{(i,j) | s_i = \bar{s}_j\}|$ be the number of matching pairs in $s$ and $\bar{s}$. This step is based on an exact algorithm by Apostolico and Guerra \cite{hunt1977fast} to solve \lcs\ in time $\tildorder(n+R)$. The main idea behind the algorithm of Apostolico and Guerra \cite{hunt1977fast} is to construct an array $M$ of size $R$, where each element of $M$ is a pair of indices $(a,b)$ that corresponds to a matching pair of $s_a$ and $\bar{s}_b$ ($s_a = \bar{s}_b$). The construction of $M$ also requires the pairs to be sorted, that is, pair $(a,b)$ comes before pair $(a',b')$ if and only if either $a < a'$ or $a = a'$ and $b < b'$.

It follows from the construction of $M$ that finding $\flcs$ is equivalent to finding the longest subsequence of $M$ that is increasing in terms of both the first and the second element of each pair. Apostolico and Guerra ~\cite{hunt1977fast} show that such an increasing subsequence can be found in time $\tildorder(n+R)$ using a classic data structure. Here, we slightly modify the method of ~\cite{hunt1977fast} to present a randomized $\tildorder(n+R/\alpha)$  algorithm with an approximation factor of $\alpha$. The overall idea is to sample each element of $M$ with a rate of $1/\alpha$ and solve the problem for the sampled set. Notice that since we wish to keep the running time $\tildorder(n+R/\alpha)$, we do not afford to generate array $M$ before sampling the elements since the size of $M$ may be more than $\tildorder(n+R/\alpha)$. However, we explain later in the section that the sampled array can be constructed in time $O(n+R/\alpha)$ and thus an $\alpha$ approximation of \lcs\ can be obtained in time $\tildorder(n+R/\alpha)$.

\setcounter{algocf}{4}
\begin{algorithm2e}
	\KwData{
		$s$ and $\bar{s}$. $R$: number of matching pairs of $s$ and $s'$}
	\KwResult{An $\alpha$-approximate solution for \lcs.}
	$M^* = $ empty array \;
	$j \leftarrow 0$ \;
	$P =$ a geometric probability distribution on $\{1,2,\ldots,\infty\}$ with $P(Y = k) = (1-\alpha)^{k-1} \alpha$ \; 
	$z\leftarrow 0$\;
	\While{$j<R$}{
		Sample $x$ from $P$;  \Comment{$x$: distance to the index of the next element}\\
		$j \leftarrow j+x$ \;
		\If{$j\leq R$}{
			
			$p \leftarrow$ $j$'th element of $M$; \Comment{$M$ is the set of matching pairs}\\ \Comment{We dont have access to $M$ directly, but $M[j]$ can be found in time $\tilde{O}(1)$}\\
			$M^*[z]= p$\;
			$z\leftarrow z+1$\;
		} 
	}
	Find the longest subsequence of $M^*$ which is increasing in terms of both the elements.
	\vspace{0.0cm}
	\caption{\samarr}
	\label{alg:6}
\end{algorithm2e}

Using the above algorithm, we modify the input to guarantee $m = O(n^{1/2+\eta})$ for some $\eta\geq 0$. For a given threshold $\tau$, define a symbol $c$ in string $t$ to be low frequency, if $\fr_c(t) \leq \tau$ and high frequency otherwise. For a string $t$, we define $t_\LLL$ and $t_\HHH$ to be subsequences of $t$ subject to only low and only high frequency symbols, respectively. For instance, if $t = \mathsf{aabccd}$ and $\tau = 1$ we have $t_\LLL = \mathsf{bd}$ and $t_\HHH = \mathsf{aacc}$.

It follows from the definition that for any threshold $\tau$ we have  
\begin{align*}
 &|\llcs(s_\LLL,\bar{s}_\LLL)| + |\llcs(s_\LLL,\bar{s}_\HHH)|+\\ & |\llcs(s_\HHH,\bar{s}_\LLL)| + |\llcs(s_\HHH,\bar{s}_\HHH)| \geq |\flcs|
 \end{align*}
which in turn implies 
$$
\max \Bigg\{\begin{array}{c}|\llcs(s_\LLL,\bar{s}_\LLL)| ,\\ |\llcs(s_\LLL,\bar{s}_\HHH)| ,\\ |\llcs(s_\HHH,\bar{s}_\LLL)| ,\\ |\llcs(s_\HHH,\bar{s}_\HHH)|\end{array}\Bigg\} \geq |\flcs|/4.
$$

Hence, by solving $\lcs$ for all four combinations and returning the best solution, one can provide a $4$-approximation solution to $\flcs$. Indeed this argument also carries over to any approximate solution, that is, approximating the subproblems within a factor of $\alpha$ implies a $4 \alpha$ approximate solution for $\mathsf{lcs}(s,\bar{s})$. We show that except the subproblem $(s_\HHH,\bar{s}_\HHH)$ the number of matching pairs ($R$) for the rest of the subproblems is bounded and thus a desirable solution for such instances can be obtained using Algorithm \ref{alg:6}. Hence, the only nontrivial instance is the case of $(s_\HHH,\bar{s}_\HHH)$. Therefore, if the solution of Algorithm \ref{alg:7} is not desirable, it only suffices to find an approximate solution for $\llcs(s_\HHH,\bar{s}_\HHH)$. Moreover, since each symbol appears at least $\tau$ times in any of $s_\HHH$ and $\bar{s}_\HHH$, the number of symbols of $s_\HHH$ and $\bar{s}_\HHH$ is bounded by $n/\tau$.
\setcounter{algocf}{1}
\begin{algorithm2e}
\KwData{
$s$ and $\bar{s}$.}
\KwResult{An approximate solution for \lcs.}
$\tau \leftarrow n^{1/2-\eta}$ \; 
Construct $s_\LLL,s_\HHH,\bar{s}_\LLL,\bar{s}_\HHH$ \;
Approximate $\llcs(s_\LLL,\bar{s}_\LLL)$, $\llcs(s_\LLL,\bar{s}_\HHH)$, and $\llcs(s_\HHH,\bar{s}_\LLL)$ using Algorithm \ref{alg:6} and return the best\;
\caption{\textsf{low, high frequency algorithm}}
\label{alg:7}
\end{algorithm2e}

Algorithm \ref{alg:7} has no assumption on the input and outputs three strings $\solution,s'$ and $\sbar'$, where $s'$ and $\sbar'$ are subsequences of $s$ and $\sbar$. For a fixed $0 < \eta <1/2$, Algorithm \ref{alg:7} guarantees that either $\solution$ is an  $O(n^{1/2-\eta})$ approximation solution,  or $|\llcs(s',\sbar')|\geq |\flcs|/4$ and $|\Sigma'| = O(n^{1/2+\eta})$, where $\Sigma'$ is the set of symbols appeared in $s'$ and $\sbar'$. 

	\begin{tcolorbox}
		\textsf{Inputs:} $s$ and $\bar{s}$.\\
		\textsf{Outputs:} $\solution,s',\sbar'$, where $s'$ is a subsequence of $s$ and $\sbar'$ is a subsequence of $\sbar$.\\
		\textsf{Guarantee:}
		Either  $\solution$ is an $O(n^{1/2-\eta})$ approximation of $\llcs(s,\sbar)$, or
		\begin{itemize}
			\item $|\llcs(s',\sbar')| \geq |\flcs|/4$.
			\item $|\Sigma'| = O(n^{1/2+\eta})$.
		\end{itemize}
	\end{tcolorbox}

As mentioned earlier, the basic idea behind Algorithm \ref{alg:7} is to decompose each of $s$ and $\sbar$ into two substrings based on the frequency of the symbols and solve the subproblems corresponding to the four combinations. For this purpose, we define the concept of low and high frequency symbols. 
Let $\tau = n^{1/2-\eta}$ be a threshold and 
define a symbol $c$ to have a low frequency in string $t$, if $\fr_c(t) \leq \tau$. Otherwise, $c$ has a high frequency. In addition, let $t_{\LLL}$ and $t_{\HHH}$ be the substrings of $t$ restricted to the symbols with low and high frequencies, respectively. 
 Since $s_{\LLL}$ and $s_{\HHH}$ cover $s$ and $\bar{s}_{\LLL}$ and $\bar{s}_{\HHH}$ cover $\bar{s}$, we have
\begin{align}
&|\llcs(s_\LLL,\bar{s}_\LLL)| + |\llcs(s_\LLL,\bar{s}_\HHH)|+ \nonumber\\ & |\llcs(s_\HHH,\bar{s}_\LLL)| + |\llcs(s_\HHH,\bar{s}_\HHH)| \geq |\flcs|. \label{ineq:1}
\end{align}
By pigeonhole principle, we can imply that at least one of the four terms on the left hand side of Inequality \eqref{ineq:1}, approximates $\llcs(s,\sbar)$ within a factor of $4$. As we show in Section \ref{atfts}, the first three instances can be $O(n^{1/2-\eta})$ approximated in time $\tilde{O}(n)$. Using the method we describe in Section \ref{atfts}, Algorithm \ref{alg:7} finds an $O(n^{1/2-\eta})$ approximation solution for each of the first three subproblems and selects the largest one. Let $\solution$ be the selected solution. The output of Algorithm \ref{alg:7} consists of three strings $\solution,s_{\HHH},\sbar_{\HHH}$. If 
$$
\max \Bigg\{\begin{array}{c}|\llcs(s_\LLL,\bar{s}_\LLL)| ,\\ |\llcs(s_\LLL,\bar{s}_\HHH)| ,\\ |\llcs(s_\HHH,\bar{s}_\LLL)| ,\\ |\llcs(s_\HHH,\bar{s}_\HHH)|\end{array}\Bigg\} \geq |\flcs|/4
$$
holds, then
$\solution$ is an $O(n^{1/2-\eta})$ approximation of $\flcs$. 
Otherwise, due to Inequality \eqref{ineq:1}, we have $|\llcs(s_{\HHH},\sbar_{\HHH})| \geq |\flcs|/4$. In addition, since each symbol in $s_{\HHH}$ and $\sbar_{\HHH}$ has a frequency of at least $\tau = n^{1/2-\eta}$, the total number of different symbols in $s_{\HHH},\sbar_{\HHH}$ is at most $n/n^{1/2-\eta} = n^{1/2+\eta}.$

The result of this section is formally stated in Lemma \ref{lem:sec6}. 
\begin{lemma}
	\label{lem:sec6}
	Let $\solution,s_{\HHH},\sbar_{\HHH}$ be the output of Algorithm \ref{alg:7} for input strings $s$ and $\sbar$ and $0 \leq \eta \leq 1/2$ be a fixed number. Then, either  $|\llcs(s_{\HHH},\sbar_{\HHH})| \geq |\flcs|/4$ holds, or 
	$\solution$ is an $O(n^{1/2-\eta})$ approximation solution. Also, $s_{\HHH}$ and $\sbar_{\HHH}$ have at most $n^{1/2+\eta}$ symbols.
\end{lemma}

In the next section, we propose a random procedure to provide an $O(n^{1/2-\eta})$ approximation solution for the first three subinstances.

\subsection{Instances with Low Frequency Symbols.} \label{atfts}

 We claim that for the first three subinstances on the left-hand side of Inequality \eqref{ineq:1}, we can find an $O(n^{1/2-\eta})$ approximation solution for $\lcs$ in near linear time. A character $s_i$ matches a character $\sbar_j$, if they have the same symbols. The key property of these  subinstances is that the number of pairs that match in these subinstances is small, as we prove in Lemma \ref{mpsm}.
\begin{lemma}
	\label{mpsm}
	Let $r$ and $\bar{r}$ be two strings of size $n$, with the property that for every symbol $c$, we have $\fr_c(r) \leq \tau$ and let $$R := |\{(i,j) | r_i = \bar{r}_j\}|$$ be the number of matching pairs of $r$ and $\bar{r}$. Then, we have $R\leq \tau n$.

\end{lemma}
\begin{proof}
		Let $f_i$ be the frequency of the $i$'th character of $\bar{r}$ in $r$. The total number of the matching pairs of
	$r$ and $\bar{r}$ is $$\sum_{1 \leq i \leq |\bar{r}|} f_i.$$ Since $r$ contains only symbols with frequency at most $\tau$, 
	$f_i \leq \tau$ holds for every $1\leq i \leq |r|$. Furthermore, we have $|\bar{r}| = n$, which implies
	\begin{align*}
	\sum_{i \leq |r|} f_i &\leq \tau n .
	\end{align*}
	
\end{proof}

Since each of the instances $(s_{\LLL},\bar{s}_{\LLL}), (s_{\LLL},\bar{s}_{\HHH}) ,$ and  $(s_{\HHH},\bar{s}_{\LLL})$  contains at least one string with low frequency characters, the total number of the matching pairs in each instance is upper bounded by  $n^{3/2-\eta}$. Now, we show that we can find an $O(n^{1/2-\eta})$ approximation solution for these instances in time $\tilde{O}(n)$.  First, we start by stating that a near-linear time algorithm can exactly compute $\lcs$ in time $\tilde O(n+R)$.
 
\begin{theorem}[Apostolico and Guerra \cite{apostolico1987longest}]
 	\label{clm:1}
 Let $r$ and $\bar{r}$ be two strings of size $n$, and let $$R := |\{(i,j) | r_i = \bar{r}_j\}|$$ be the number of matching pairs in $r$ and $\bar{r}$. Then, we can compute $\llcs(r,\bar{r})$ in time $\tilde{O}(n+R)$.
 \end{theorem} 
\begin{proof}
Let $M$ be an array of size $R$, where each element  of $M$ is a pair of indices that corresponds to a matching pair. In addition, suppose that the elements in $M$ are sorted according to the following rule:
\begin{equation}
\label{compare}
\small (i,j)<(i',j') \iff i<i' \mbox{ or } (i=i' \mbox{ and } j<j').
\end{equation}

It can be easily observed that finding the longest common subsequence of $s$ and $\sbar$ is equivalent to finding the longest subsequence of $M$ that is increasing in terms of both the elements of each pair, which can be found in time $O(n+R\log R) = \tilde{O}(n+R)$ \cite{hunt1977fast}.	
\end{proof}

Based on the method proposed in Theorem \ref{clm:1}, we present Algorithm \ref{alg:6} which is an $\tildorder(n+R/\alpha)$ time algorithm with an approximation factor of $\alpha$. Suppose that the input to Algorithm \ref{alg:6} are two strings $r,\bar{r}$ of sizes $n$ and let $R$ be the number of matches in $(r,\bar{r})$. Recall the definition of $M$ in the proof of Theorem \ref{clm:1}. Algorithm \ref{alg:6} constructs an array $M^*$  where every pair of $M$  is in $M^*$ with probability $1/\alpha$. The expected size of $M^*$ is $R/\alpha$ and since each pair appears in $M^*$ with probability $1/\alpha$, the expected size of the longest increasing subsequence of $M^*$ is at least $|\llcs(r,\bar{r})|/\alpha$ which approximates$\llcs(r,\bar{r})$ with a factor of $\alpha$.  Notice that, when $R=O(n^{3/2-\eta})$, choosing $\alpha = n^{1/2-\eta}$ results in  expected running time $\tilde{O}(n)$ and expected approximation factor $n^{1/2-\eta}$.

The only challenge here is that the size of $M$ is large, and we cannot construct $M$ before constructing $M^*$. We argue that we can construct $M^*$ in time $\tilde{O}(R/\alpha)$ from $r$ and $\bar{r}$. To do so, we use Observation \ref{obs:ran}.

\begin{observation}
	\label{obs:ran}
Let $P$ be a probability distribution over $\{0,1\}$, with $P(0)=\alpha$ and $P(1)=1-\alpha$, and let $x_1,x_2,\ldots,x_n$ be $n$ random variables sampled independently from $P$. For every $i$, define $next(i)$ as the minimum index $j$, such that $j>i$ and $x_j=1$. Then, the distribution of $next(i)$ for every $i$ is a geometric distribution with parameter $\alpha$. 
\end{observation}

Based on Observation \ref{obs:ran}, we iteratively sample the index of the next element of $M^*$ in $M$ using a geometric distribution with parameter $\alpha$ which determines its distance to the currently selected index.  
Next, for every sampled number $i$, we find the $i$'th matching pair of $M$. Note that to find the $i$'th matching pair of $M$, we do not need to explicitly access $M$; we can find it in amortized time $O(1)$ using a data structure of size $O(n)$, which for every symbol $c$ stores an array $\bar{F}_c$ of size $\fr_c(\bar{r})$ containing the occurrence indices of $c$ in $\sbar$. More precisely, suppose that the last selected pair is $(i,j)$ which is the $k$'th matching pair in $M$ and $next(k) = k'$. Then, we find the least index $i' \geq i$, such that 
	$$ \textsf{rem}(j) + \sum_{i<l\leq i'} \fr_{s_l}(\bar r) \geq k'-k  $$
where $\textsf{rem}(j)$ is the number of indices $j'>j$ with $\sbar_{j'} = \sbar_j$, which can be computed using $\bar{F}_{\sbar_j}$:
$$
\textsf{rem(j)} = |\bar{F}_{\sbar_j}| - l
$$
where $l$ is the index of $\bar{F}_{\sbar_j}$ that $\bar{F}_{\sbar_j} = j$. After finding $i'$, using $\bar{F}_{s_{i'}}$ we can find the pair $(i',j')$ which is the $k'$'th pair.


\begin{lemma}
	\label{lem:matchlow}
	Let $r$ and $\bar{r}$ be two strings of size $n$, with the property that the number of matching pairs of $s$ and $\sbar$ is $O(n^{3/2-\eta})$. Then, by setting $\alpha=R/n$ Algorithm \ref{alg:6} returns an $O(n^{1/2-\eta})$ approximation of $\flcs$ in time ${O}(n)$.
\end{lemma}
\begin{proof}
	The expected length of the longest increasing subsequence of $M^*$ is at least $|\flcs|/2\alpha\geq |\flcs|/2n^{1/2-\eta} $. Hence, Algorithm \ref{alg:6} outputs an $O(n^{1/2-\eta})$ approximation of $\flcs$ in linear time. 
\end{proof}


%% file: Boundingsumfreq.tex
\section{Algorithms \ref{alg:3} and \ref{alg:4}: Decomposition into Blocks and Permutations}\label{bbrs}

The last step of our algorithm is technically more involved as we perform two separate algorithms and argue that the better of the two algorithms gives us a desirable solution for a class of inputs. 

\begin{figure*}	
	\begin{tcolorbox}
		\begin{minipage}{0.3\textwidth}
			Algorithm \ref{alg:3}:\\[0.5cm]
			\textsf{Inputs:} $s$ and $\bar{s}$.\\
			\textsf{Precondition}
			\begin{itemize}
				\item $|\flcs| \geq n^{1-\delta}$
				\item $m\leq n^{1/2+\eta}$
			\end{itemize}
		\end{minipage}
		\hfill\vline\vline\hfill
		\begin{minipage}{0.4\textwidth}
			Algorithm \ref{alg:4}:\\[0.5cm]
			\textsf{Inputs:} $s$ and $\bar{s}$.\\
			\textsf{Precondition}
			\begin{itemize}
				\item $|\flcs| \geq n^{1-\delta}$
				\item $m\leq n^{1/2+\eta}$
			\end{itemize}
		\end{minipage}
		\\[1cm]
		The approximation factor of the better solution returned by Algorithms \ref{alg:3} and \ref{alg:4}:  $$\tilde{O}(\brruntime).$$
	\end{tcolorbox}
\end{figure*}	

Formally, during this section, we analyze Algorithm \ref{combine}, which takes $s$ and $\sbar$ as input and runs both Algorithms \ref{alg:3} and \ref{alg:4} on the input and returns the better solution. 
Algorithm \ref{combine} assumes that the input satisfies two conditions: first, the solution size is at least  $n^{1-\delta}$, and second, the number of different symbols in $s$ and $\sbar$ is at most $n^{1/2 + \eta}$, i.e., $|\Sigma|\leq n^{1/2+\eta}$. In this section, we first analyze Algorithm \ref{alg:3}, and characterize the conditions under which Algorithm \ref{alg:3} provides the desirable approximation guarantee. Next, we show that the inputs that do not satisfy these conditions can be approximated well  by Algorithm \ref{alg:4}. The main argument of this section is formally stated in Lemma \ref{lem:alg8}.

\begin{lemma}
	\label{lem:alg8}
	For two strings $s$ and $\sbar$ with the properties that $|\flcs| \geq n^{1-\delta}$ and $|\Sigma| \leq n^{1/2+\eta}$, Algorithm \ref{combine} returns an $\tilde{O}(\brruntime)$ approximation solution for $\flcs$.
\end{lemma}

Throughout this section, we use Observation \ref{dbd} several times. Observation \ref{dbd} is a simple consequence of Pigeonhole principle.

\begin{observation}
	\label{dbd}
	Let $X = \{x_1,x_2,\ldots,x_{|X|}\}$ be a set of  numbers with the property that $\sum_i x_i = S$ and $\max_i x_i = M.$ then, at least $S/M$ of the numbers in $X$ have value at least $S/|X|$.   
\end{observation}

\setcounter{algocf}{5}
\begin{algorithm2e}[t]
	\KwData{
		$s$ and $\bar{s}$.}
	\KwResult{An approximate solution for \lcs.}	
	$\textsf{lcs}_1 =$ the solution returned by Algorithm \ref{alg:3}\;
	$\textsf{lcs}_2 =$ the solution returned by Algorithm \ref{alg:4}\;
	return $\arg \max \{|\llcs_1|, |\llcs_2|\}$ 
	\caption{block-to-block + random shifting Algorithm}
	\label{combine}
\end{algorithm2e}

\subsection{Algorithm \ref{alg:3}: Block to Block Algorithm}

We start by analyzing Algorithm \ref{alg:3}. Algorithm \ref{alg:3} divides $s$ and $\bar{s}$ into $\sqrt n$ blocks of size $\sqrt n$. Denote by $\block_i$ and $\barblock_i$, the $i$'th block of $s$ and $\bar{s}$, respectively. Algorithm \ref{alg:3} constructs a $\sqrt{n} \times \sqrt{n}$ table $T$ as follows: for every $i$ and $j$, Algorithm \ref{alg:3} chooses a random position in $\block_i$. Let $c$ be the symbol in the randomly selected position. We set $$T[i][j] = \min \{ \fr_c(\block_i), \fr_c(\barblock_j)\}.$$   
For brevity, denote by $c_{i,j}$ the symbol selected for blocks $b_i$ and $\barblock_j$. Obviously, we can fill table $T$ in time $O(n)$. Using $T$, Algorithm \ref{alg:6} finds the longest common subsequence of $s$ and $\bar{s}$ with the following two constraints:

\begin{itemize}  
	\item 
	No two characters in the same block of $s$ are matched to the characters in different blocks of $\sbar$ and vice versa.  
	
	\item If a block $\block_i$ is matched to a block $\barblock_{j}$, then all the matched symbols of $\block_i$ and $\barblock_{j}$ are $c_{i,j}$.
\end{itemize}

Denote by $\clcs(s,\bar{s})$, the longest common subsequence of $s$ and $\bar{s}$ with the constraints described above. We can use dynamic programming to find $\clcs(s,\bar{s})$ as follows. Let $D$ be a $\sqrt{n} \times \sqrt{n}$ table where $D[i][j]$ stores the length of the longest restricted common subsequence of the strings corresponding to the first $i$ blocks of $s$ and the first $j$ blocks of $\bar{s}$. Then, $D[i][j]$ can be computed via the following recursive formula:
\begin{equation}
\label{eq:2}
D[i][j] \hspace{-1mm}= \hspace{-1mm}\begin{cases}
0 & \hspace{-4mm} i\mbox{ or } j=0 \\
\max \Bigg\{ \begin{array}{c} D[i][j-1],\\D[i-1][j],\\ T[i][j]+ D[i-1][j-1]\end{array} \Bigg\} &i,j>0
\end{cases}
\end{equation}
The total running time of Algorithm \ref{alg:3} is $O(n)$. 
We now determine the conditions under which Algorithm \ref{alg:3} returns an $O(n^{1/2-\zeta})$ approximation solution for a fixed $\zeta$. For this, we first in Lemma \ref{lem:block_match} show that there exists a set of $\Omega(n^{1/2-2\delta})$ pairs of blocks, where each block pair contributes $\Omega(n^{1/2-\delta})$ characters to $\flcs$.
\begin{lemma}
	\label{lem:block_match}
	There exist two sequences $i_1 < i_2 < \ldots < i_{l}$ and $j_1 < j_2 < \ldots < j_{l}$ of indices such that $l \geq n^{1/2-2\delta}/8$, and for every $k$, $|\llcs(\block_{i_k},\barblock_{j_k})|\geq  n^{1/2-\delta}/2.$ 
\end{lemma}
\begin{proof}
	Let $\alpha_i$ and $\beta_i$ be the block numbers that the $i$'th character of  $\flcs$ belongs to in $s$ and $\bar{s}$, respectively, and let $U$ be a multi-set containing  pairs $(\alpha_i,\beta_i)$. We know both the sequences $\alpha_1,\alpha_2, \ldots, \alpha_{|\flcs|}$ and $\beta_1,\beta_2, \ldots, \beta_{|\flcs|}$ are non-decreasing. Thus, if for some $j$, $(\alpha_j,\beta_j) \neq (\alpha_{j+1},\beta_{j+1})$, at least one of the following two inequalities holds: $$\alpha_{j+1} >\alpha_j \qquad \mbox{ or }\qquad \beta_{j+1}>\beta_j.$$ As a result, the total number of different pairs in $U$ is at most $2\sqrt{n}$, and since the size of a block is $\sqrt n$, each pair appears no more than $\sqrt n$ times in $U$. 	Since $|U| \geq n^{1-\delta}$, by Observation \ref{dbd} we can find at least  $n^{1/2-\delta}$ different pairs which appear at least $n^{1/2-\delta}/2$ times in $U$. Let $\widetilde{U}$ be the set of these pairs. A pair $(\alpha_i,\beta_i)$ collides with a pair $(\alpha_j,\beta_j)$, if $\alpha_i = \alpha_j$ or $\beta_i = \beta_j$. Since the size of each block in $s$ and $\bar{s}$ is $\sqrt n$ and each pair of $\widetilde{U}$ appears at least  $n^{1/2-\delta}/2$ times in $U$, no pair $(\alpha_i,\beta_i)$ in $\widetilde{U}$ can collide with more than $4n^\delta$ other pairs in $\widetilde{U}$. Otherwise, at least one of $\block_{\alpha_i}$ or $\barblock_{\beta_i}$ must have a size more than $n^{1/2}$.
	Hence, $\widetilde{U}$ contains at least $n^{1/2-2\delta}/8$ collision-free pairs. 
	Each one of these pairs corresponds to a pair of blocks in $s$ and $\bar{s}$ with the desired property.
\end{proof}

Let 
$
\widetilde{U}
$ be the set of $l$ block pairs mentioned in Lemma \ref{lem:block_match}. Now, we partition these block pairs into two types. 	A symbol $c$ is crebris in block pair $(\block_{i},\barblock_{j})$, if $$\min \{\fr_c(\block_i),\fr_c(\barblock_j)\}\geq n^{4\delta + \zeta},$$
and is rarus, otherwise. A pair $(\block_{i},\barblock_{j}) \in \widetilde{U}$ approximately consists of crebris symbols, if 
	$$
	\sum_{c \in C}\fr_c(\block_{i})\geq n^{1/2-2\delta}/4, 
	$$
where, $C$ is the set of crebris symbols of $(\block_i,\barblock_j)$. The intuition behind this definition  is as follows: if a pair  $(\block_{i},\barblock_{j})$ approximately consists of crebris characters, when we select a random positions of $\block_{i}$ to fill $T[i][j]$, with probability at least  $$n^{1/2-2\delta}/4n^{1/2} = n^{-2\delta}/4,$$ we have $T[i][j] \geq n^{4\delta + \zeta}$. Now, let $\crebrisblocks$ be the set of pairs in $\widetilde{U}$ which are approximately consisted of crebris characters and $\rarusblocks = \widetilde{U} \setminus \crebrisblocks$. 

\begin{lemma}
	\label{lemover}
	If $|\crebrisblocks| \geq |\widetilde{U}|/2$, then the expected length of the solution returned by Algorithm \ref{alg:3} is  $\Omega(n^{1/2+\zeta})$.
\end{lemma}
\begin{proof}
As mentioned above, for every block pair $(\block_i,\barblock_j) \in \crebrisblocks$, with probability at least $n^{-2\delta}/4$ we have $T[i][j] \geq n^{4\delta + \zeta}$, which means that the expected value of $T[i][j]$ is at least $n^{2\delta + \zeta}/4$. Let $t$ be the solution returned by Algorithm \ref{alg:3}. Since Algorithm \ref{alg:3} selects the best constrained solution, we have 
\begin{align*}
\mathbb{E}(t) &\geq  \sum_{(\block_{i},\barblock_{j})\in B} \mathbb{E}(T[i][j])\\
&\geq  \sum_{(\block_{i},\barblock_{j})\in \crebrisblocks} \mathbb{E}(T[i][j])\\
&\geq n^{1/2-2\delta}/16 \cdot n^{2\delta + \zeta}/4\\
&=  n^{1/2+\zeta}/64\\
&=\Omega(n^{1/2+\zeta}).
\end{align*}
\end{proof}



In the next section, we show that for the case that $|\rarusblocks|\geq |\widetilde{U}|/2$, Algorithm \ref{alg:4} provides a good approximate solution. 

\subsection{Algorithm \ref{alg:4}: Random Shifting Algorithm.} \label{sec:rndshift}

In this section, we analyze Algorithm \ref{alg:4} and  show that this algorithm performs well on the instances that satisfy the preconditions, and have the property that $|\rarusblocks| \geq |\widetilde{U}|/2$.
Let us first give a short description of  Algorithm \ref{alg:4}. Algorithm \ref{alg:4}  modifies each block of the strings into a special structure, which we call semi-permutation. Next, Algorithm \ref{alg:4} samples a random integer $r$ uniformly from $[1,\sqrt n]$, and for every $1\leq i \leq \sqrt n$, finds $q_i = \llcs(\block_i,\barblock_{(i+r) \mod \sqrt n})$
\footnote{Here we need a slightly modified $\mod$ operator where $\sqrt n \mod \sqrt n = \sqrt n$.}
. Finally, Algorithm \ref{alg:4} selects one of $$t_1 = q_1q_2\ldots q_{\sqrt n-r}$$ or $$t_2 = q_{\sqrt n-r+1}q_{\sqrt n-r+2}\ldots q_{\sqrt n}$$ which is longer, as an approximation of $\flcs$. In what follows we describe Algorithm \ref{alg:4} in more details. 

In the first step, Algorithm \ref{alg:4} updates $s$ and $\bar{s}$ by performing the following operation on each block: for every symbol which appears more than once in that block, we keep one of the instances uniformly at random and remove the rest. After this operation, each symbol  appears in each block at most once. Note that, the blocks may have different sizes as some symbols might not appear in some blocks. Formally, after this operation, each block of $s$  and $\sbar$ is a permutation of the characters in $\Sigma$, except that some of the symbols are missing in each block. We call such a structure \emph{semi-permutation}.

As we show in Section \ref{semip}, after this operation, there exist two sets $U$ and $\bar{U}$ of the blocks in $s$ and $\sbar$, such that $|U| = |\bar{U}| = \Omega(\Usize)$, and for every two blocks
$\block_i$ and $\barblock_{j}$ that are randomly selected from $U$ and $\bar{U}$,  we have


\begin{equation}
\label{expsizeran}
\mathbb{E}(|\llcs(\block_{i},\barblock_{j})|) = 
\tilde{\Omega}(\RandUsize).
\end{equation}
In the second step, Algorithm \ref{alg:4} selects a random integer $r$ and for every $i$, finds  $$q_i = \llcs(\block_i,\barblock_{(i+r) \mod \sqrt n}).$$
Since both $\block_i$ and $\barblock_j$ are semi-permutations, they have at most $\sqrt n$ matching pairs and hence, $q_i$ can be computed in time $\tilde{O}(\sqrt n)$ \cite{apostolico1987longest}.  Therefore, the running time of Algorithm \ref{alg:4} is $\tilde{O}(n)$. 
To analyze the approximation factor of Algorithm \ref{alg:4}, we focus on the blocks in $U$ and $\bar{U}$.  Note that since $r$ is selected uniformly at random, if for some index $i$, both $\block_i \in U$ and $\barblock_{(i+r) \mod \sqrt n} \in \bar{U}$ hold, then we have 
$$
\mathbb{E}(|q_i|) = \tilde{\Omega}(\RandUsize).
$$
Furthermore, since $|U|=|\bar{U}|= \Omega(\Usize)$ and the total number of blocks is $n^{1/2}$,  the expected number of such indices is at least
$
 |U|\cdot |U|/n^{1/2}   = \Omega(n^{1/2-52\delta-8\zeta-2\eta}).
$ Thus, the expected size of the solution returned by Algorithm \ref{alg:4} is at least 
\begin{align*}
(\sum_{1\leq i \leq \sqrt n} \mathbb{E}(|q_i|))/2 = 
\tilde{\Omega}(n^{2/3-221/3\delta-34/3\zeta-10/3\eta}). 
\end{align*}
Since $|\flcs|\leq n$, the approximation factor of Algorithm \ref{alg:4} is
$$
\tilde{\Omega}(n^{1/6+221/3\delta+34/3\zeta+10/3\eta}). 
$$

Finally, we know that the approximation factor of Algorithm \ref{alg:3} is $O(n^{1/2-\zeta})$. To optimize the better solution of Algorithms \ref{alg:3} and \ref{alg:4}, we set
\begin{align*}
1/6+221/3\delta+34/3\zeta+10/3\eta &=
 1/2-\zeta \\
 1/2+221\delta+34\zeta+10\eta &= 3/2 - 3\zeta \\
\end{align*}
which implies
$$
\zeta = 1/37 - (221/37)\delta-(10/37)\eta.\\
$$
 Therefore, the better solution returned by these two algorithms is an $$\tilde{O}(\brruntime)$$ approximation. 
 

The remaining challenge is to prove the existence of $U$ and $\bar{U}$, 
 which we prove in the next section.

\subsubsection{Rarus Blocks and Semi-permutations}
\label{semip}
We call a block rarus, if it belongs to a pair in $\rarusblocks$.
Recall the property of the block pairs in $\rarusblocks$: for each pair $(\block_i,\barblock_{j}) \in \rarusblocks$, the total number of the crebris characters
\footnote{We use rarus and crebris for both symbols and characters. Rarus (crib) characters are those whose corresponding symbols are rarus (crebris).}
 (i.e., a character that its corresponding symbol has frequency at least $n^{4\delta +\zeta} $ in both $\block_i$ and $\barblock_j$) in $\block_{i}$ is at most $n^{1/2-2\delta}/4$. The rest of the characters in $\block_i$ are rarus. Since $|\llcs(\block_{i},\barblock_j)|\geq n^{1/2-\delta}/2$, at least  
\begin{equation}
\label{raruscont}
n^{1/2-\delta}/2 - n^{1/2-2\delta}/4 \geq
n^{1/2-\delta}/4
\end{equation}
of the characters in $\llcs(\block_i,\barblock_j)$ are rarus. By definition, rarus characters either have a frequency at most $n^{4\delta +\zeta}$ in both the blocks, or have a frequency less than $n^{4\delta +\zeta}$ in one block and at least $n^{4\delta +\zeta}$ in the other one. We argue that at least half of the rarus characters have a frequency at most $\tau = 16n^{5\delta+\zeta}$ in both $\block_{i}$ and $\barblock_{j}$. To prove this, consider the symbols with  frequency more than $\tau$ in $\block_i$ and $\barblock_j$. Since $|\block_i|,|\barblock_j|\leq n^{1/2}$, there are at most $2n^{1/2}/(16n^{5\delta +\zeta})$ different symbols of this type in $\block_i$ and $\barblock_j$. The contribution of each of these symbols to $\llcs(\block_i,\barblock_{j})$ is at most $n^{4\delta+\zeta}$. Thus, the total contribution of these symbols to $\llcs(\block_i,\barblock_{j})$ is at most $n^{1/2-\delta}/8.$
By Inequality \eqref{raruscont}, the total contribution of rarus characters to $\llcs(\block_i,\barblock_j)$ is at least $n^{1/2 - \delta}/4$. Therefore, at least 
$$n^{1/2-\delta}/4 - n^{1/2-\delta}/8 = 
n^{1/2-\delta}/8
$$
of the characters in $\llcs(\block_i,\barblock_j)$ correspond to the rarus characters with frequency at most $\tau$ in both $\block_i$ and $\barblock_j$. 

Consider a character $c$ in $\llcs(\block_{i},\barblock_{j})$ where its corresponding characters in $s$ and $\sbar$ are rarus and have frequency at most $\tau$. In the first step of Algorithm \ref{alg:4}, we select one instance of each symbol in each block. Thus, with probability at least $1/\tau^2$, Algorithm \ref{alg:4} retains the characters that correspond to $c$ in $s$ and $\sbar$. This implies that the expected size of $\llcs(\block_i,\barblock_j)$ after the first step is at least 
$
n^{1/2-\delta}/(8\tau^2). 
$
Furthermore, after the first step, we have  
\begin{align*}
\mathbb{E}(|\llcs(s,\sbar)|) &\geq |\rarusblocks| \cdot n^{1/2-\delta}/(8\tau^2) \\
&\geq n^{1-3\delta}/(64\tau^2).
\end{align*}

In addition, each symbol appears at most once in each block of $s$ and $\bar{s}$.  In other words, each block of $s$ and $\sbar$ is a permutation of a subset of the symbols in $\Sigma$. 
Finally, after the refinement, we further restrict our attention to the symbols that appear in at least 
$n^{1/2-3\delta -\eta}/(64\tau^2)$
of the blocks. Note that 
since $
\mathbb{E}(|\llcs(s,\sbar)|) \geq 
 n^{1-3\delta}/(64\tau^2)
$, we have at most $n^{1/2}$ blocks, and the number of the symbols is at most $n^{1/2+\eta}$, by Observation \ref{dbd} we  conclude that there are at least 
$n^{1/2-3\delta}/(64\tau^2)$
 symbols, which appear in at least 
 $n^{1/2-3\delta-\eta}/(64\tau^2)$
  blocks. Let $\widetilde{\Sigma}$ be the set of these symbols. Thus, after the refinement process, the probability that each of the symbols in $\widetilde{\Sigma}$ appears in a random block is at least 
$$\alpha = n^{1/2-3\delta-\eta}/(n^{1/2}64\tau^2) = \Omega(n^{-13\delta-2\zeta-\eta}).$$\
 We denote this value by $\alpha$. The contribution of the symbols in $\widetilde{\Sigma}$ to $\flcs$ is  $\Omega(n^{1-26\delta -4\zeta-\eta}).$ Again, using Observation \ref{dbd} we can imply that there are $\Omega(n^{1/2-26\delta -4\zeta-\eta})$ of the block pairs in $\rarusblocks$, each of which contributing $$ \Omega(n^{1/2-26\delta -4\zeta-\eta})$$ characters with symbols in $\widetilde{\Sigma}$  to $\flcs$. We define $U$ and $\bar{U}$ as the set of these blocks in $s$ and $\sbar$, respectively. Furthermore, for each block in $U$ and $\bar{U}$, we only consider the characters that their corresponding symbols are in $\widetilde{\Sigma}$ and ignore the rest. Thus, we assume that the blocks in $U$ and $\bar{U}$ are semi-permutations with symbols in $\widetilde{\Sigma}$.

In the rest of the section, we prove a lower bound on the expected size of the longest common subsequence of two random blocks of $U$ and $\bar{U}$. To prove the lower bound, we first provide a method to convert a semi-permutation into a complete permutation, without a significant increase in the expected size of the $\lcs$ of two random blocks. Next, we prove a lower bound on the expected size of two random complete permutations which in turn imposes a lower bound on the expected size of two semi-permutations in $U$ and $\bar{U}$.
 






We describe the method to convert the semi-permutations in $U$ to complete permutations. The same procedure can be applied to the semi-permutations in $\bar{U}$. The process of converting the blocks in $U$ to  complete permutations is formally stated in Algorithm \ref{refine}. In the beginning, Algorithm \ref{refine} constructs a complete permutation, namely $\mask{s}$, by appending the characters in $\log n /\alpha$ blocks in $\rarusblocks$ as follows: initially, $\mask{s}$ is empty. To convert $\mask{s}$ to a complete permutation, Algorithm \ref{refine} chooses $\log n/\alpha$ blocks of $s$ that belong to the pairs in $\rarusblocks$ uniformly at random. After selecting the blocks,  Algorithm \ref{refine} iterates over these blocks one by one, and for each block $b$, appends the symbols of $\widetilde{\Sigma}$ that are in $b$, but not in $\mask{s}$ to the end of $\mask{s}$, with the same order as they appear in $b$. Finally, there might be still some symbols in $\widetilde{\Sigma}$ that have not appeared in $\mask{s}$. If so, we add them to the end of $\mask{s}$ with an arbitrary order.
After constructing $\mask{s}$, we convert every semi-permutation $\block_i \in U$ to a complete permutation by appending the symbols in $\mask{s}$ that are not appeared in $\block_{i}$ to the end of $\block_{i}$ with the same order as they appeared in $\mask{s}$. 

\setcounter{algocf}{6}
\begin{algorithm2e}
	\KwData{
		$U$}
	\KwResult{$U'$: a set of complete permutations.}
	$\mask{s} = \emptyset $\;
	$B:$ a set of $\log n/\alpha$  blocks of $s$ that belong to  $\rarusblocks$, selected uniformly at random\;
	\For{each $b \in B$}{
		$\tilde{b} = $ maximal substring of $b$ only consisted of the characters in $\widetilde{\Sigma} \setminus \mask{s}$ \;
		$\mask{s} \leftarrow \mask{s} + \tilde{b}$; \Comment{$\tilde{b}$ is appended to the end of $\mask{s}$}
	}
	\If {$\mask{s}$ is not a permutation of $\widetilde{\Sigma}$}{
		Add the characters in $\widetilde{\Sigma} \setminus \mask{s}$ with an arbitrary order to the end of $\mask{s}$
	}
	\For{each $b_i \in U$}{
		$b'_i = b_i$ \; 
		$b = $ maximal substring of $\mask{s}$ only consisted of the characters in $\widetilde{\Sigma} \setminus \tilde{b'_i}$ \;
		$b'_i = b'_i + b$ \;
	}
	return $U' = \{b'_1,b'_2,\ldots,b'_{|U|}\}$
	\caption{\textsf{semi-permutation to complete-permutation}}
	\label{refine}
\end{algorithm2e}
 
Let $U'$ and $\bar{U}'$ be two sets of blocks, containing the complete permutations produced by executing Algorithm \ref{refine} on $U$ and $\bar{U}$. We now prove Lemma \ref{lem} for $U'$ and $\bar{U}'$. 

\begin{lemma}
	\label{lem}
	Let $\beta$ be the expected length of the longest common subsequence of two random blocks of $U$ and $\bar{U}$ and  $\block'_i,\barblock'_j$ be two random blocks of $U'$ and $\bar{U}'$. Then, the expected size of $\llcs(\block'_i,\barblock'_j)$ is  $O(\beta \log n /\alpha)$.  
\end{lemma} 	

\begin{proof}
	First, note that $\mask{s}$  is consisted of two parts: the part that is appended to $\mask{s}$ with respect to the random blocks of $\rarusblocks$, and the part that includes the symbols which are added to $\mask{s}$ in an arbitrary order. We denote these two parts by $w$ and $a$, respectively (and $\bar{w},\bar{a}$ for $\mask{\bar{s}}$). Recall that the probability that each symbol appears in each  random block is $\alpha$. Considering this, we claim that the expected size of  $a$ is $O(1)$. 
	Let $p_c$ be the probability that after appending the extra symbols in $\log n/\alpha$ randomly selected blocks, $\block'_i$ does not contain symbol $c$. By definition, $c$ appears in each of the randomly selected blocks with probability at least $\alpha$. Hence, after $\log n/\alpha$ steps, we have 
	$$
	p_c \leq (1-\alpha)^{\log n/\alpha} \simeq e^{-\log n} \leq 1/n.
	$$
	On the other hand, we know that the number of the characters in $\Sigma$ is $O(n)$ which concludes that the expected number of the characters in $a$ is $O(n\cdot p_c) = O(1)$. The same argument also implies $\bar{a} = O(1)$.
	Furthermore, each of $w$ and $\bar{w}$ is a concatenation of the characters in $\log n/\alpha$ random blocks.
	
	Now, consider two blocks $\block'_i$ and $\barblock'_j$ selected uniformly at random from $ U'$ and $\bar{U}'$. Note that we can argue that each of $\block'_i$ and $\barblock'_j $ is a concatenation of the characters in at most $\log n / \alpha$ random blocks plus a string of size $O(1)$.  With an argument similar to the one used in the proof of Lemma \ref{lem:block_match}, we can imply that the number of different block pairs that contribute to $\llcs(\block'_i,\barblock'_j)$ is
	$O(\log n/\alpha)$. Furthermore, we assume that the expected size of the longest common subsequence of two random blocks of $U$ and $\bar U$ is $\beta$, which 
	concludes that the expected size of the longest common subsequence of $\block'_i$ and $\barblock'_j$ is $O(\beta \log n/\alpha)$. 
\end{proof}

We now prove Lemma \ref{lowerbd} which provides a lower bound on the expected size of the longest common subsequence of two random blocks of $U'$ and $\bar{U}'$. 

\begin{lemma}
	\label{lowerbd}
	The expected size of the longest common subsequence of two randomly selected blocks of $U'$ and $\bar{U}'$ is $\Omega(n^{(1/2-26\delta -4\zeta-\eta)/3})$. 
\end{lemma}

As mentioned in Section \ref{overview}, to prove Lemma \ref{lowerbd}, we use two well-known theorems, one in the context of order theory and the other in the  graph theory. Here we present these two theorems and use them to prove Lemma \ref{lowerbd}. We begin by Dilworth's theorem. Before we state the theorem, we need some basic definitions.

A partially ordered set (poset) $P$ is consisted of a ground set $X$ and a partial order $\preceq$ that satisfies the following conditions:
\begin{itemize}
	\item (reflexive) $a \preceq a$ for all $a \in X$.
	\item (assymetricity) If $a \preceq b$ and $b \preceq a$, then we have $a = b$.
	\item (transitivity) If $a \preceq b$ and $b \preceq c$, then we have $a \preceq c$.
\end{itemize}
A chain of length $k$ is defined as a sequence $a_1,a_2,\ldots,a_k$ of elements in $X$, such that $a_1 \preceq a_2 \preceq \ldots \preceq a_k$. An antichain of size $k$ is defined as a set $\{a_1,a_2,\ldots,a_k\}$, where for every $i,j \in [k]$, $a_i$ and $a_j$ are incomparable, i.e., neither $a_i \preceq a_j$ nor $a_j \preceq a_i$ holds. A chain decomposition of $X$ is a partition of the elements of $X$ into disjoint chains. 

\vspace{0.2cm}
{\noindent T{\footnotesize HEOREM} (Dilworth-1950 \cite{dilworth1950decomposition}). \textit{    Let $\ell$ be the size of the maximum antichain in poset $P$. Then, $P$ can be decomposed into $\ell$ chains.}}
\vspace{0.2cm}

Here, we use the dual form of this theorem, which is proved by Mirsky in 1971 \cite{mirsky1971dual}.

\begin{thm}[dual of Dilworth \cite{mirsky1971dual}]
	Let $\ell$ be the size of the maximum chain in poset $P$. Then, $P$ can be decomposed into $\ell$ disjoint antichains.  
\end{thm}

Now, let $\pi$ and $\bar \pi$ be two permutations of the characters in $\Sigma$. According to $\pi$ and $\bar \pi$, we define a poset $P$ as follows: the ground set is $\Sigma$, and for two symbols $c_1,c_2 \in \Sigma$, we have $c_1 \preceq c_2$, if and only if $c_1$ appears before $c_2$ in both $\pi$ and $\bar \pi$. It is easy to observe that any chain in $P$ is corresponding to a common subsequence of $\pi$ and $\bar{\pi}$.  Therefore, we can restate the dual form of Dilworth's theorem as follows. 
\begin{theorem}[Dilworth-restated]
	\label{dilworth}
	Given two strings $\pi$ and $\bar{\pi}$, each of which is a permutation of the symbols in  $\Sigma$, where $|\Sigma| = m$ . If $|\llcs(\pi,\bar{\pi})| = x$, then $\bar{\pi}$ can be decomposed into $x$ substrings, such that the characters in each of these strings appear exactly in the reverse order as appeared in $\pi$, i.e., the longest common subsequence of each of these substrings and $\pi$ is $1$.
\end{theorem}
We use a nontrivial consequence of Theorem \ref{dilworth} which we state in Lemma \ref{cor:1}.
\begin{lemma}
	\label{cor:1}
	\label{lem:3block}
	Let $\pi_1, \pi_2, \pi_3$ be three permutations of the symbols in $\Sigma$ where $|\Sigma| = m$. Then, 
	$$
	|\llcs(\pi_1,\pi_2)| \cdot |\llcs(\pi_2,\pi_3)| \cdot |\llcs(\pi_3,\pi_1)| \geq m.
	$$
\end{lemma}

\begin{proof}
	Consider $\pi_1$ and $\pi_2$ and suppose that $|\llcs(\pi_1,\pi_2)| = x$ and $|\llcs(\pi_1,\pi_3)|=y$. According to Theorem \ref{dilworth}, $\pi_2$ and $\pi_3$ can be respectively decomposed into $x$ and $y$ substrings, such that the characters in each of these substrings appear in the reverse order as appear in $\pi_1$. On the other hand, by the Pigeonhole principle, at least one of the substrings of $\pi_1$ and one of the substrings of $\pi_2$ have $m/xy$ characters in common. Since these characters appear in the same order in both of these substrings, the longest common subsequence of these two substrings is at least $m/xy$. Therefore
	\begin{align*}
	|\llcs(\pi_1,\pi_2)| \cdot |\llcs(\pi_2,\pi_3)| \cdot |\llcs(\pi_3,\pi_1)| &\geq  x \cdot y \cdot m/xy \\
	&= m.
	\end{align*}	
\end{proof}

The second theorem we use in the proof of  Lemma \ref{lowerbd} is called \emph{Tur\'an}, which provides a  lower bound on the edge size of  $K_{r+1}$-free graphs. In Section \ref{overview} we mentioned a general form of this theorem. By setting $r=2$ in Tur\'an's theorem we obtain the following Corollary. 

\begin{corollary}[of Tur\'an's theorem]	\label{turan}
	Any triangle-free graph on $k$ vertices has at most $k^2/4$ edges.
\end{corollary}

Now, we are ready to prove Lemma \ref{lowerbd}

\begin{proof}[of Lemma \ref{lowerbd}]
	Recall that the blocks in $U$ and $\bar{U}$ constitute $|U|$ block pairs, namely $(\block_1,\barblock_1),(\block_2,\barblock_2),\ldots,(\block_{|U|},\barblock_{|U|})$, such that for every $1\leq i \leq |U|$, we have
	$$
	\llcs(\block_i,\barblock_i) =  \Omega(n^{1/2-26\delta -4\zeta-\eta}).
	$$ 
	With this in mind, we construct a graph on the blocks of $U'$ and $\bar{U}'$. Let 
	$$
	d = \min_{1\leq i \leq |U|} |\llcs(\block_i,\barblock_i)|,
	$$
	 and let $G(V,E)$ be a graph, where $|V|=|U|$ and there is an edge between $v_{i}$ and $v_{j}$, if at least one of the following two conditions holds:
	\begin{itemize}
		\item $\llcs(\block'_{i},\barblock'_{j}) \geq \sqrt[3]{d}$.
		\item $\llcs(\block'_{j},\barblock'_{i}) \geq \sqrt[3]{d}$.
	\end{itemize}
	
	Now, consider tree vertices $v_i,v_j,v_k \in V$. By definition, we know $|\llcs(\block'_{i},\barblock'_i)| \geq d$. If we only consider these $d$ symbols in blocks $\block'_j$ and $\barblock'_k$, 
	by Lemma \ref{cor:1}, we have
	\begin{align*}
	|\llcs(\block'_j,\barblock'_k)| \cdot |\llcs(\block'_j,\llcs(\block'_{i},\barblock'_i))| \cdot |\llcs(\llcs(\block'_{i},\barblock'_i),\barblock'_k)| &\geq  d, 
	\end{align*}
	which means
	\begin{align*}
	\max \Bigg\{\begin{array}{c}|\llcs(\block'_j,\barblock'_k)|\\ |\llcs(\block'_j,\llcs(\block'_{i},\barblock'_i))| \\ |\llcs(\llcs(\block'_{i},\barblock'_i),\barblock'_k)|\end{array}\Bigg\}&\geq \sqrt[3]{d}.
	\end{align*}
	
	Therefore, at least one edge exists between every triple $v_i,v_j$, and $v_k$, and hence the complement graph of $G$ (i.e., $\bar{G}$) is triangle-free. By Corollary \ref{turan}, $\bar{G}$ has at most $|U|^2/4$ edges, and so $G$ has at least $$|U|(|U|-1)/2 - |U|^2/4 = |U|^2/4 - |U|/2$$ edges. For $|U| \geq 4$, this value is at least $|U|^2/8$. Thus, 
	
	\begin{align*}
	\sum_{i,j} |\llcs(\block'_{i},\barblock'_{j})| \geq d^{1/3}(U^2/8).  
	\end{align*}
	Since there are $|U|^2$ pairs of blocks, the expected size of the longest common subsequence of two randomly selected blocks of $U'$ and $\bar{U}'$ is at least 
	$$\Omega(d^{1/3}) =  \Omega(n^{(1/2-26\delta -4\zeta-\eta)/3}).$$	
\end{proof}

Finally, the combination of Lemmas \ref{lem} and \ref{lowerbd} implies that the expected length of the longest common subsequence of two random blocks of $U$ and $\bar U$ is 
$$
\Omega(n^{(1/2-26\delta -4\zeta-\eta)/3} \cdot n^{-13\delta-2\zeta-\eta}/\log n),
$$
that is 
$$
\tilde{\Omega}(n^{1/6-65/3\delta-10/3\zeta-4/3\eta}).
$$




%% file: acknowledgement.tex
\section{Acknowledgment}
The third author would like to thank Alexandr Andoni for helpful discussion.